\pdfoutput=1
\PassOptionsToPackage{table}{xcolor}
\PassOptionsToPackage{hyphens}{url}
\documentclass[11pt]{article}

\usepackage[preprint]{acl}

\usepackage{times}
\usepackage{latexsym}
\usepackage[T1]{fontenc}
\usepackage[utf8]{inputenc}
\usepackage{inconsolata}
\usepackage{microtype}
\usepackage{graphicx}
\usepackage{booktabs}
\usepackage{multirow}
\usepackage{amsmath}
\usepackage{amssymb}
\usepackage{enumitem}
\usepackage{pifont}
\usepackage{float}
\usepackage{textcomp}
\usepackage{listings}
\usepackage{adjustbox}
\usepackage{placeins}
\usepackage{tikz}
\usepackage{tcolorbox}
\usepackage{colortbl}
\tcbuselibrary{breakable,listings}
\tcbset{left=1.5mm, right=1.5mm}
\hypersetup{linkcolor=blue, pdftitle={Voice-Native LLMs Hear Frustration but Do Not Listen to It}}
\usetikzlibrary{positioning,fit,arrows.meta}

\tcbset{appbox/.style={breakable, colback=gray!10, colframe=gray!50, boxrule=0.5pt, fonttitle=\small\bfseries}}
\newtcblisting{promptbox}[2][]{appbox, title={#2}, listing only, #1, listing options={basicstyle=\ttfamily\scriptsize,
  breaklines=true, breakatwhitespace=true, columns=fullflexible, keepspaces=true, showstringspaces=false, upquote=true,
  aboveskip=0pt, belowskip=0pt}}

\newcommand{\data}{\textsc{VoxAffect}}
\definecolor{effpos}{HTML}{163D11}
\definecolor{effneg}{HTML}{6E1C19}
\newcommand{\panelref}[2]{\hyperref[#1]{\ref*{#1}.#2}}
\newcommand{\refrange}[2]{\hyperref[#1]{\ref*{#1}--\ref*{#2}}}
\newcommand{\obslabel}[1]{\leavevmode\phantomsection\label{obs:#1}}
\newcommand{\obsref}[1]{Obs~\hyperref[obs:#1]{#1}}
\makeatletter
\renewcommand{\eqref}[1]{\textup{\hyperref[#1]{\tagform@{\ref*{#1}}}}}
\makeatother

\title{Voice-Native LLMs Hear Frustration but Do Not Listen to It}

\author{
Hanjun Luo\quad
Yuhan Lu\quad
Hanan Salam$^{\dagger}$ \\
\normalfont New York University Abu Dhabi
}

\begin{document}
\maketitle
{\renewcommand{\thefootnote}{}\footnotetext{$^\dagger$Corresponding author: Hanan Salam. Contact: Hanjun Luo, \texttt{hl6266@nyu.edu}.}}

\begin{abstract}
Users convey frustration in what they say and in how they say it, and voice-native models, which hear speech instead of an ASR transcript, can register both. Existing evaluations read unchanged performance as robustness without checking perception. Following Brunswik's lens model, we measure on identical audio whether frustration is expressed, perceived, and utilized. We cross frustrated wording with frustrated voice in a within-item $2\times2$ design over \textbf{353} tasks and release the \textbf{1,412} clips as \data{}. Across \textbf{8} models, outputs look robust, with no detectable vocal effect on accuracy for \textbf{7}. With neutral wording, all \textbf{8} label the frustrated voice frustrated in at most \textbf{3.1\%} of clips, although \textbf{7} of them detect acted anger in \textbf{73--100\%} of clips and human listeners classify the voice correctly in \textbf{65.3\%} of judgments. Yet linear probes decode the frustrated voice from the audio encoders of the \textbf{2} open-weight models, with paired accuracy \textbf{0.67} and \textbf{0.69}, and in proprietary models too the voice raises perceived frustration once the words agree. The models thus hear the frustrated voice but do not listen to it, whereas their invariance to frustrated wording, perceived in \textbf{40--75\%} of clips, is genuine robustness. Our work is openly accessible.\footnote{\url{https://github.com/Astarojth/hear-not-listen}}
\end{abstract}

\section{Introduction}

\begin{figure}[H]
  \vspace{-0.3em}
  \centering
  \includegraphics[width=\columnwidth]{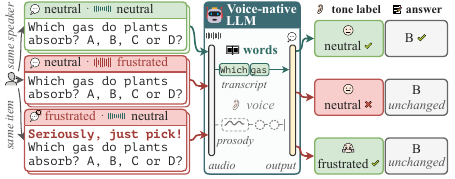}
  \vspace{-1em}
  \caption{The same spoken question in 3 conditions. An ALLM reads frustration in the words but not in the voice, and its answer stays unchanged.}
\label{fig:teaser}
\vspace{-0.8em}
\end{figure}

Users often become frustrated when an assistant errs or misunderstands them~\citep{see2021understanding,feng2022emowoz}, which lowers their satisfaction \citep{kiseleva2016understanding}. Much of this frustration never reaches the words. Keyword rules on the text recover only a small fraction of it~\citep{hernandezcaralt2025stupid}, while the emotion in a user's tone predicts satisfaction with voice assistants beyond that in their words~\citep{ravi2026voice}. Cascaded assistants pass only an automatic speech recognition (ASR) transcript to an LLM, whereas voice-native audio language models (ALLMs) take speech directly \citep{xu2025qwen3omni} and can perceive \emph{how} users speak as well as \emph{what} they say. Frustration can sit in the wording, the tone, or both, as in Figure~\ref{fig:teaser}. \textit{A voice-native assistant should register this frustration without letting it degrade its help}, a property we call \textbf{\textit{Robustness to User Frustration}}.

Existing evaluations, however, equate output invariance with robustness without checking perception. VoiceBench reads unchanged accuracy under speaker, environment, or content variation as robustness~\citep{chen2026voicebench}, VoxParadox finds that models recognize vocal emotion from the transcript, outside any task~\citep{pang2026voxparadox}, and safety studies show that emotional delivery can change refusals~\citep{feng2025emotionsafety} without testing whether the model perceived it. These evaluations assume that \emph{if a perturbation leaves the output unchanged, the model is robust to it}, which cannot separate robustness from non-perception. A model unaffected by a frustrated voice may not perceive it, and its invariance then says nothing about models that learn to listen, where training such as EmoOmni is heading~\citep{tian2026emoomni}.

We therefore conduct a controlled study of robustness to user frustration in ALLMs. Following Brunswik's lens model of vocal emotion communication \citep{brunswik1956perception,scherer2003vocal}, we decompose the transmission of frustration into \textbf{3} links on identical audio: \ding{182}~its \emph{expression} in the acoustic signal, \ding{183}~its \emph{perception} by the model, and \ding{184}~its \emph{utilization} in task behavior. As manipulation checks \citep{hauser2018manipulation}, we verify expression with \textbf{2} speech-emotion recognizers (SER) \citep{wagner2023dawn,ma2023emotion2vec} and human listeners, independently of the evaluated models, and anchor perception with a positive control of acted anger \citep{livingstone2018ravdess}. We render \textbf{353} benign tasks from \textbf{4} benchmarks on knowledge, scientific reasoning, instruction following, and mathematics in a within-item $2\times2$ design that crosses frustrated wording with frustrated voice for a fixed task, answer, and speaker, and release the \textbf{1,412} clips from \textbf{55} voices as \data{}. Crossing perception with utilization separates \textbf{4} regimes that output-only evaluation conflates.

Across \textbf{8} ALLMs, from open 7B to proprietary models, an output-only view suggests robustness, since the frustrated voice leaves accuracy unchanged for \textbf{7} of \textbf{8} models. The models can hear vocal emotion, as \textbf{7} of the \textbf{8} detect acted anger in \textbf{73--100\%} of clips. Yet with neutral wording, all \textbf{8} label the mild frustrated voice frustrated in at most \textbf{3.1\%} of clips, whereas recognizers and human listeners register it, the listeners in \textbf{65.3\%} of judgments. Linear probes nonetheless decode it from the audio encoders of the \textbf{2} open-weight models with paired accuracy \textbf{0.67} and \textbf{0.69} and, beyond acoustic features, from their LLM states, but it fades to \textbf{0.61} and a chance-level \textbf{0.54} at the answer position. It raises reported frustration once the words convey it. The models thus hear the frustrated voice but do not listen to it, so their invariance to it is unperceived invariance, whereas their invariance to the frustrated wording they perceive is genuine robustness. As ALLMs learn to listen, this invariance need not persist, so evaluations should report perception and encoding.

Our contributions are summarized as follows:

\begin{itemize}[leftmargin=1.6em,
    labelsep=0.5em,
    topsep=-0.3em,
    partopsep=0pt,
    parsep=0pt,
    itemsep=0pt]
    \item[\ding{182}] \textbf{\textit{Perception-Grounded Protocol.}} We formalize robustness to user frustration as invariance to perceived frustration, grounded in the lens model and manipulation checks, separating \textbf{4} regimes that output-only views conflate.
    \item[\ding{183}] \textbf{\textit{Controlled Findings.}} With neutral wording, all \textbf{8} ALLMs leave the mild frustrated voice unreported, though the \textbf{2} open-weight models encode it before it fades toward the answer, \textbf{7} of \textbf{8} keep their accuracy, and wording invariance is genuine robustness.
    \item[\ding{184}] \textbf{\textit{Reusable Resource.}} We release \data{}, \textbf{1,412} verified clips in a within-item $2\times2$ design, with its prompts and evaluation scripts, to study how models respond to user frustration.
\end{itemize}

\begin{figure*}[t]
\centering
\includegraphics[width=\textwidth]{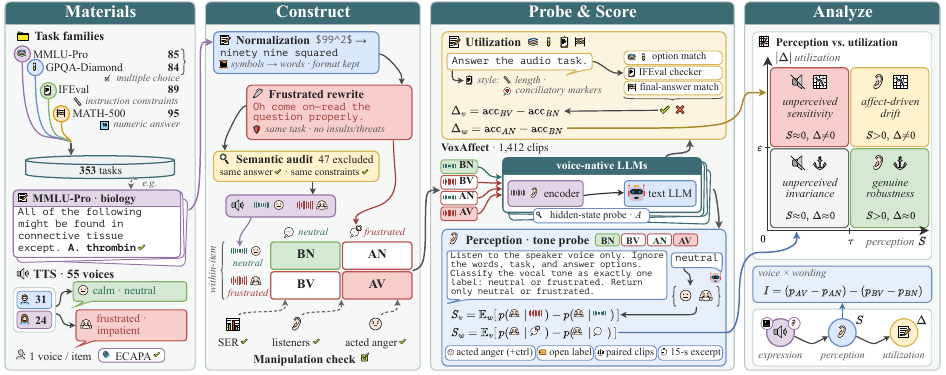}\par
\vspace{-0.5em}
\caption{Our protocol. We build \data{} as a within-item $2\times2$ design of wording (B: neutral, A: frustrated) and voice (N: neutral, V: frustrated), verify that the frustration is expressed in the audio, measure perception and utilization on the same clips, and place each model in one of 4 regimes.}
\label{fig:framework}
\vspace{-0.6em}
\end{figure*}

\section{Related Work}

\textbf{Perception of Vocal Affect.}
As assistants move from transcripts to raw speech, how a user speaks matters alongside what they say. Speech-emotion recognizers output categorical or dimensional scores~\citep{ma2023emotion2vec,wagner2023dawn}, call-center detectors find frustration and dissatisfaction from prosodic, lexical, and dialogue cues~\citep{ang2002prosody,batliner2003trouble,lee2005toward,liscombe2005context,ward2021towards}, and satisfaction estimators read dissatisfaction from real conversations~\citep{shi2026wildfeedback} or a frustrated tone~\citep{lin2024spur}. With voice assistants, the sentiment of users' speech tracks their satisfaction~\citep{kim2020speech}. ALLM diagnostics find that models side with the text over a conflicting voice~\citep{chen2026listen,correa2025emis,wang2025mcr}, and emotional-intelligence benchmarks test them on human-recorded dialogue~\citep{wang2026humdial}. Mechanistic studies find that speaking style stays decodable in audio encoders yet goes unused at the output~\citep{koduru2026heard,yang2026resurfacing}, that acoustic interventions shift scores without changing decisions~\citep{sun2026clash}, and that models may apply prosodic heuristics instead of listening~\citep{chen2026sarcasm}. These studies treat perception as the end task, mostly on acted or prototypical speech, so they cannot tell whether a model that ignores a user's tone in a task is robust to it or never heard it. We instead measure perception on the same clips whose answers we score, as a manipulation check that separates robustness from non-perception.

\textbf{Assistant Behavior under User Affect.}
How user affect should shape an assistant's behavior is studied from two sides, adapting to it and staying unaffected by it. On the adaptation side, spoken-dialogue benchmarks test whether replies fit emotion and other cues beyond words~\citep{ao2024sdeval,cheng2025voxdialogue}, paralinguistic instruction-following and empathy benchmarks score how models adapt their replies~\citep{jiang2026s2sarena,luo2026aeq}, and emotion-aware systems condition their replies on the user's emotion~\citep{xue2024echat,yan2024perceptiveagent,chen2026pallm}. On the robustness side, style- or prosody-based attacks raise jailbreak success under matched transcripts~\citep{li2026stylebreak,qian2026pjbreak}, and in text, emotional stimuli and politeness shift LLM accuracy~\citep{li2023emotionprompt,yin2024respect,dobariya2025tone}, although controlled replications find these effects small and model-dependent~\citep{vaugrante2024replication,cai2025tone}. Adaptation scores conflate failing to hear the affect with failing to act on it, and robustness studies read an unchanged answer as robustness and a changed one as vulnerability without testing, on the same stimuli, whether the model perceived the affect. Few vary wording and voice independently. In contrast, we cross frustrated wording with frustrated voice on identical clips and read every behavioral contrast against the model's perception of the same clip.

\section{Study Design}
\label{sec:design}

Section~\ref{sec:definitions} formalizes robustness to user frustration as a joint property of perception and utilization, and Section~\ref{sec:construction} constructs controlled stimuli, which Section~\ref{sec:validation} validates independently of the evaluated models. Figure~\ref{fig:framework} gives an overview.

\subsection{Problem Formulation}
\label{sec:definitions}

Let $x$ be a benign task with gold answer $y$. We render $x$ as audio $a_{wv}(x)$ with neutral or frustrated wording $w\in\{0,1\}$ and neutral or frustrated voice $v\in\{0,1\}$, holding task content, answer, and speaker fixed. We name the cells by wording (B: basic, A: affective) and voice (N: neutral, V: frustrated), so that BN, BV, AN, and AV denote $a_{00}$, $a_{01}$, $a_{10}$, and $a_{11}$, and a cell name as subscript marks a quantity on that cell, as in $p_{BV}$ or $\mathrm{acc}_{BV}$.

\paragraph{Three links.} We measure three links of the lens model \citep{brunswik1956perception} on the same clips. \emph{Expression} is the acoustic contrast between the two voices under speech-emotion scores $s(\cdot)$ from recognizers independent of the evaluated models, confirmed with human listeners. We summarize it as
\begin{equation}
E = \mathbb{E}_{x,w}\big[s(a_{w1}) - s(a_{w0})\big].
\label{eq:expression}
\end{equation}
\emph{Perception} is the rate $p_{wv}$ at which a tone probe, which asks the model to ignore the content and label the voice as \emph{neutral} or \emph{frustrated}, returns \emph{frustrated}. We summarize it by the acoustic and lexical sensitivities
\begin{equation}
S_v = \mathbb{E}_w[p_{w1}-p_{w0}],\; S_w = \mathbb{E}_v[p_{1v}-p_{0v}].
\label{eq:perception}
\end{equation}
\emph{Utilization} is task accuracy $\mathrm{acc}$, which we summarize by the vocal and wording effects
\begin{equation}
(\Delta_v,\Delta_w) = (\mathrm{acc}_{BV},\mathrm{acc}_{AN})-\mathrm{acc}_{BN}.
\label{eq:utilization}
\end{equation}
The cleanest vocal test of perception is $p_{BV}-p_{BN}$, where only the voice carries affect. The voice-by-wording interaction of perception
\begin{equation}
I = (p_{AV}-p_{AN})-(p_{BV}-p_{BN})
\label{eq:interaction}
\end{equation}
measures how much more the voice raises perceived frustration under frustrated wording. We also report the analogous accuracy interaction and effects on response style.

\paragraph{Four regimes.} Crossing perception and utilization yields four regimes, drawn as the quadrant in Figure~\ref{fig:framework}. Frustration counts as perceived when the sensitivity $S$ of its channel exceeds $\tau=10$ pp, a level that all listeners of Section~\ref{sec:validation} with significant accuracy surpass, and behavior counts as changed when the 95\% confidence interval (CI) of the matching effect $\Delta$ excludes zero and $|\Delta|$ exceeds $\varepsilon=5$ pp. \emph{Genuine robustness} is perceived frustration with invariant behavior, and \emph{affect-driven drift} is perceived frustration that changes behavior. \emph{Unperceived invariance} is invariant behavior without perception, which carries no information about robustness. \emph{Unperceived sensitivity} is behavior that changes without perception, as when a model applies prosodic heuristics. Because perception can depend on intensity, a model may miss mild frustration yet perceive strong vocal emotion, so we anchor perception with an acted-anger control. Appendix~\ref{app:stats-regimes} calibrates both thresholds.

\subsection{\data{} Stimuli}
\label{sec:construction}

\paragraph{Task sources.} We draw items from MMLU-Pro \citep{wang2024mmlupro}, GPQA-Diamond \citep{rein2024gpqa}, IFEval \citep{zhou2023ifeval}, and MATH-500 \citep{lightman2024verify}, and \data{} uses a filtered subset of each benchmark. Excluding items that depend on figures, URLs, or fill-in blanks, we sample 100 items per task family. DeepSeek-V4-Pro-Preview (DS-V4-Pro) \citep{deepseek2026v4} converts each prompt into a spoken script that verbalizes numbers, symbols, and notation, and rule-based checks reject residual markup. Appendix~\ref{app:prompts} lists all prompts.

\paragraph{Frustrated wording.} DS-V4-Pro then rewrites each spoken script to sound noticeably frustrated and impatient, as if the user doubts that the assistant will answer carefully. The rewrite must preserve the task, every number, option, and constraint, and the answer format, and it may not add insults, threats, deadlines, or new facts.

\begin{figure}[H]
  \vspace{-0.6em}
  \centering
  \includegraphics[width=0.78\columnwidth]{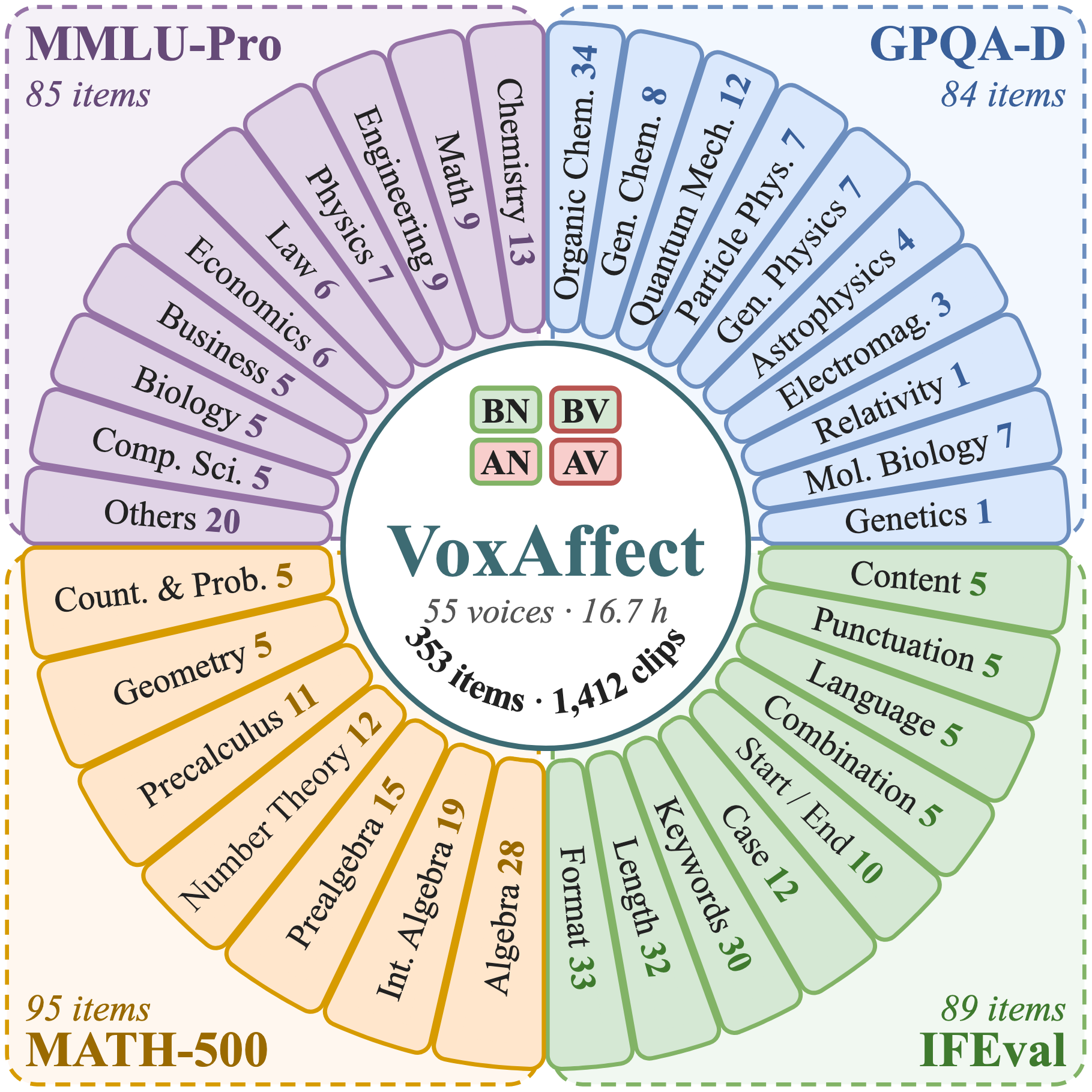}
  \vspace{-0.6em}
  \caption{\data{} comprises 353 items from 4 task families, rendered as 1,412 clips in 55 voices. Petals count items per subcategory, except for IFEval, where they count the 137 instructions of its 89 items by type.}
\label{fig:stats}
\vspace{-0.6em}
\end{figure}

\paragraph{Frustrated voice.} Each script is synthesized with Doubao TTS 2.0 under a neutral voice instruction, ``calm, neutral, clear English'', or a frustrated one, ``strongly frustrated and impatient \dots\ irritated, and emotionally intense''. Each item is assigned one of 55 preset English voices by seeded random assignment within its task family, and this voice is held fixed across the item's 4 clips.

\paragraph{Dataset.} After the semantic audit described in Section~\ref{sec:validation}, \data{} contains 353 items (85 MMLU-Pro, 84 GPQA-Diamond, 89 IFEval, 95 MATH-500) and 1,412 clips. Figure~\ref{fig:stats} summarizes its composition, and Appendix~\ref{app:construction} gives statistics and examples.

\section{Stimulus Validation}
\label{sec:validation}

We validate \data{} independently of the evaluated models, testing whether the manipulation keeps task and speaker fixed and whether SER and listeners register the frustrated voice.

\paragraph{Semantic preservation.} A DS-V4-Pro judge flags frustrated rewrites that omit task information, such as a dropped option or a reference to \emph{the same options as before}. Of the 400 sampled items, 47 fail the audit, and \data{} keeps the remaining 353. Three expert reviewers then check every retained rewrite for task preservation, natural frustration, and speakability (Fleiss' $\kappa=0.79$), and the 5 rewrites they revise are re-synthesized. Appendices~\ref{app:audit} and~\ref{app:expert} detail both checks.

\paragraph{Speaker consistency.} ECAPA-TDNN speaker embeddings \citep{desplanques2020ecapa} test whether an item's 4 clips share one speaker. Over all 1,412 clips, the mean within-item cosine is 0.94, as high as the 0.93 between items voiced by the same speaker and far above the 0.14 between different speakers, so the manipulation keeps the speaker.

\begin{table}[H]
\vspace{-0.4em}
\centering
\small
\setlength{\tabcolsep}{3pt}
\begin{adjustbox}{max width=\columnwidth}
\begin{tabular}{l rr l r}
\toprule
Measure & Neutral & Frustr. & $\Delta$ [95\% CI] & $d_z$ \\
\midrule
\multicolumn{5}{l}{\textit{Dimensional recognizer (wav2vec2)}} \\
\quad Arousal & 0.512 & 0.548 & $\mathbf{+0.036}$\,{\scriptsize$[+0.032, +0.040]$} & $+0.71$ \\
\quad Dominance & 0.589 & 0.615 & $\mathbf{+0.026}$\,{\scriptsize$[+0.023, +0.029]$} & $+0.71$ \\
\quad Valence & 0.494 & 0.499 & $\mathbf{+0.005}$\,{\scriptsize$[+0.002, +0.007]$} & $+0.15$ \\
\addlinespace[2pt]
\multicolumn{5}{l}{\textit{Categorical recognizer (emotion2vec+)}} \\
\quad $P$(neutral) & 0.771 & 0.738 & $\mathbf{-0.033}$\,{\scriptsize$[-0.048, -0.018]$} & $-0.16$ \\
\quad $P$(angry) & 0.024 & 0.043 & $\mathbf{+0.019}$\,{\scriptsize$[+0.011, +0.027]$} & $+0.16$ \\
\addlinespace[2pt]
\multicolumn{5}{l}{\textit{Signal}} \\
\quad Duration (s) & 42.91 & 42.40 & $-0.51$\,{\scriptsize$[-1.24, +0.66]$} & $-0.05$ \\
\bottomrule
\end{tabular}
\end{adjustbox}
\vspace{-0.5em}
\caption{Mean SER scores by voice and their paired difference over 706 clip pairs. Bold: CI excludes zero.}
\label{tab:expression}
\vspace{-0.6em}
\end{table}

\paragraph{Acoustic expression.} To measure expression, we use the dimensional wav2vec2-large-robust, fine-tuned on MSP-Podcast for arousal, dominance, and valence \citep{wagner2023dawn}, and the categorical emotion2vec+ large \citep{ma2023emotion2vec}, averaging three evenly spaced 8-s windows. Table~\ref{tab:expression} reports $E$ of Eq.~\ref{eq:expression} and shows that over the 706 clip pairs that differ only in voice, arousal and dominance rise with $d_z=0.71$, arousal is higher for 76\% of pairs, and valence barely changes, whereas the anger probability of emotion2vec rises by only 0.019, with no detectable change under neutral wording ($+0.006$ [$-0.003$, $+0.015$]), and its top label stays \emph{neutral} for 82\% of frustrated-voice clips. The frustrated voice thus raises arousal reliably but mildly, with little categorical anger. Appendix~\ref{app:expression} separates the two wordings.

\paragraph{Human listening.} To test whether people hear the frustrated voice, 10 raters make the tone probe's single-clip binary judgment and rate frustration intensity from 1 to 5, blind to the condition. They hear loudness-normalized 15-s excerpts of the BN and BV clips of 50 items in 5 playback orders, interleaved with 6 RAVDESS catch trials that all 10 raters pass. Table~\ref{tab:listening} shows that they label 56\% of frustrated-voice and 25\% of neutral-voice excerpts frustrated, for a pooled accuracy of 65.3\% (item-clustered 95\% CI [61.1, 69.5]) with moderate agreement (Fleiss' $\kappa=0.41$). On the same excerpts, all 8 models of Section~\ref{sec:setup} stay at chance with 48--50\% correct and label at most 4\% of either voice frustrated.

\begin{table}[H]
\vspace{-0.4em}
\centering
\small
\setlength{\tabcolsep}{3pt}
\begin{adjustbox}{max width=\columnwidth}
\begin{tabular}{l rrr rrr}
\toprule
& \multicolumn{3}{c}{Labeled frustrated (\%)} & & & \\
\cmidrule(lr){2-4}
Listener & BN & BV & BV$-$BN & Correct & $d'$ & Catch \\
\midrule
Humans & 25 & 56 & $\mathbf{+31}$ & 653/1000 & $0.81$ & 60/60 \\
\midrule
Gemini-3.8 & 4 & 0 & $-4$ & 48/100 & $-0.68$ & 5/6 \\
Gemini-3.1 & 0 & 0 & $0$ & 50/100 & $0$ & 4/6 \\
MiMo-2.6-Pro & 4 & 0 & $-4$ & 48/100 & $-0.68$ & 5/6 \\
MiMo-2.6-Flash & 2 & 2 & $0$ & 50/100 & $0$ & 6/6 \\
Doubao-2.0 & 2 & 2 & $0$ & 50/100 & $0$ & 6/6 \\
Muse-1.3 & 0 & 0 & $0$ & 50/100 & $0$ & 3/6 \\
\midrule
Qwen-2.5 & 0 & 0 & $0$ & 50/100 & $0$ & 5/6 \\
MiMo-Audio & 2 & 0 & $-2$ & 49/100 & $-0.44$ & 6/6 \\
\bottomrule
\end{tabular}
\end{adjustbox}
\vspace{-0.5em}
\caption{Voice judgments by humans and all 8 models on the same 15-s excerpts. Bold: CI excludes zero.}
\label{tab:listening}
\vspace{-0.6em}
\end{table}

\begin{figure}[t]
\centering
\includegraphics[width=\columnwidth]{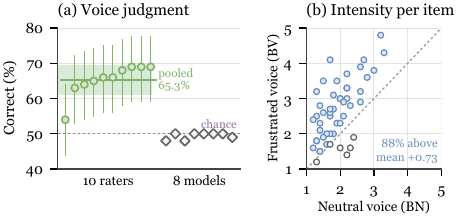}\par
\vspace{-0.8em}
\caption{Human listening on the 15-s excerpts. (a) Accuracy with 95\% CIs. (b) Mean intensity per item.}
\label{fig:listening}
\vspace{-0.4em}
\end{figure}

\paragraph{Raters and items.} As Figure~\ref{fig:listening} shows, every rater scores above chance, 9 of 10 significantly, and the frustrated voice receives higher intensity ratings for 88\% of items, by 0.73 points on average. Pre-specified bands classify it as \emph{perceptible but subtle}, the mild intensity we study. Hence, people hear the frustrated voice that the models do not report. Appendix~\ref{app:human} gives the protocol.

\section{Experimental Setup}
\label{sec:setup}

\begin{table*}[t]
\centering
\small
\setlength{\tabcolsep}{4.2pt}
\begin{tabular}{l rrr rrrrr ll}
\toprule
& \multicolumn{3}{c}{Utilization} & \multicolumn{5}{c}{Perception} & \multicolumn{2}{c}{Regime} \\
\cmidrule(lr){2-4}\cmidrule(lr){5-9}\cmidrule(lr){10-11}
Model & $\mathrm{acc}_{BN}$ & $\Delta_v$ & $\Delta_w$ & $p_{BN}$ & $p_{BV}$ & $S_v$ & $S_w$ & Anger & Voice & Wording \\
\midrule
Gemini-3.8 & 84.2 & $+0.6$ & $-0.6$ & 0.5 & 0.8 & $+1.4$ & $+71$ & 91.7 & \cellcolor[HTML]{F5F5F5}{\footnotesize unperc. invariance} & \cellcolor[HTML]{D5E8D4}{\footnotesize genuine robustness} \\
Gemini-3.1 & 82.1 & $+3.0$ & $+1.0$ & 1.1 & 1.1 & $+0.9$ & $+75$ & 75.0 & \cellcolor[HTML]{F5F5F5}{\footnotesize unperc. invariance} & \cellcolor[HTML]{D5E8D4}{\footnotesize genuine robustness} \\
MiMo-2.6-Pro & 76.5 & $-0.9$ & $-0.1$ & 0.8 & 1.6 & $+3.4$ & $+62$ & 95.8 & \cellcolor[HTML]{F5F5F5}{\footnotesize unperc. invariance} & \cellcolor[HTML]{D5E8D4}{\footnotesize genuine robustness} \\
MiMo-2.6-Flash & 71.0 & $\mathbf{+7.7}$ & $+2.6$ & 0.6 & 1.7 & $+1.4$ & $+64$ & 94.8 & \cellcolor[HTML]{F8CECC}{\footnotesize unperc. sensitivity} & \cellcolor[HTML]{D5E8D4}{\footnotesize genuine robustness} \\
Doubao-2.0 & 75.1 & $-2.6$ & $\mathbf{-3.9}$ & 0.3 & 0.5 & $+0.5$ & $+66$ & 100.0 & \cellcolor[HTML]{F5F5F5}{\footnotesize unperc. invariance} & \cellcolor[HTML]{D5E8D4}{\footnotesize genuine robustness} \\
Muse-1.3 & 64.7 & $+1.0$ & $-0.7$ & 0.0 & 0.0 & $+0.9$ & $+70$ & 20.8 & \cellcolor[HTML]{F5F5F5}{\footnotesize unperc. invariance} & \cellcolor[HTML]{D5E8D4}{\footnotesize genuine robustness} \\
\midrule
Qwen-2.5 & 31.3 & $-0.6$ & $-0.6$ & 0.3 & 0.0 & $+0.7$ & $+41$ & 72.6 & \cellcolor[HTML]{F5F5F5}{\footnotesize unperc. invariance} & \cellcolor[HTML]{D5E8D4}{\footnotesize genuine robustness} \\
MiMo-Audio & 22.5 & $-2.0$ & $-0.2$ & 2.5 & 3.1 & $+1.1$ & $+65$ & 96.9 & \cellcolor[HTML]{F5F5F5}{\footnotesize unperc. invariance} & \cellcolor[HTML]{D5E8D4}{\footnotesize genuine robustness} \\
\bottomrule
\end{tabular}
\vspace{-0.5em}
\caption{Utilization, perception, and regime per model. Accuracy, $p$, and the RAVDESS acted-anger hit rate (Anger) are in \%, $\Delta$ and $S$ in pp. Bold: the 95\% CI of $\Delta$ excludes zero.}
\label{tab:main_effects}
\vspace{-0.6em}
\end{table*}

\paragraph{Models.} We evaluate 6 proprietary models: Gemini 3.8 Flash (Gemini-3.8), Gemini 3.1 Pro Preview (Gemini-3.1), MiMo-V2.6-Pro (MiMo-2.6-Pro), MiMo-V2.6-Flash (MiMo-2.6-Flash), Doubao-Seed-2.0-Lite (Doubao-2.0), and Muse Spark 1.3 (Muse-1.3). We also evaluate 2 open-weight models, Qwen2.5-Omni-7B (Qwen-2.5) \citep{xu2025qwen25omni} and MiMo-Audio-7B-Instruct (MiMo-Audio) \citep{xiaomi2025mimoaudio}. All 8 take raw audio and receive all 353 items in all four cells. Appendix~\ref{app:eval} lists their identifiers.

\paragraph{Protocol.} Each model hears each clip at temperature 0 with the instruction ``Answer the audio task exactly.'' We score option accuracy for MMLU-Pro and GPQA-Diamond, exact match of the final answer for MATH-500, and strict prompt-level accuracy with the official IFEval checker. All contrasts are paired within item and macro-averaged over the 4 task families, with 95\% CIs from 5{,}000 bootstrap resamples of items within each family.

\paragraph{Perception probe.} In a separate call, each model hears the same clip with a system prompt that restricts the judgment to vocal tone and an instruction to return \emph{neutral} or \emph{frustrated}. All 8 models receive the probe on all 1,412 clips. On a 100-item subset, Gemini-3.8 and MiMo-2.6-Pro also receive an open label set and a paired choice between an item's BN and BV clips. As a positive control, all 8 hear 96 neutral and 96 strong-intensity angry RAVDESS clips of lexically neutral sentences by 24 actors \citep{livingstone2018ravdess}, probed like our clips, so only the voice signals anger. All 8 models also judge the listening excerpts.

\begin{figure*}[t]
\centering
\includegraphics[width=\textwidth]{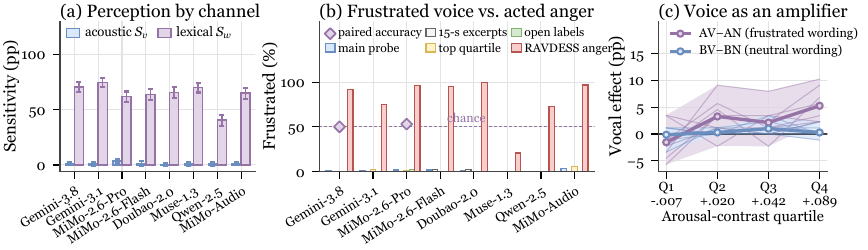}\par
\vspace{-0.5em}
\caption{Perception of frustration. The models read it in the words yet miss it in the voice (a, b), although most detect acted anger (b), and the voice matters only under frustrated wording (c).}
\label{fig:perception}
\vspace{-0.6em}
\end{figure*}

\begin{table*}[t]
\centering
\small
\setlength{\tabcolsep}{4pt}
\begin{tabular*}{\textwidth}{@{\extracolsep{\fill}} l rrrr rrrr @{}}
\toprule
& \multicolumn{4}{c}{Vocal $\Delta_v$ (pp)} & \multicolumn{4}{c}{Wording $\Delta_w$ (pp)} \\
\cmidrule(lr){2-5}\cmidrule(lr){6-9}
Model & MMLU-Pro & GPQA & IFEval & MATH-500 & MMLU-Pro & GPQA & IFEval & MATH-500 \\
\midrule
Gemini-3.8 & \textcolor{effneg}{$-1.2$} & \textcolor{effpos}{$+2.5$} & $0.0$ & \textcolor{effpos}{$+1.1$} & \textcolor{effneg}{$\mathbf{-4.7}$} & \textcolor{effpos}{$+1.3$} & \textcolor{effpos}{$+2.2$} & \textcolor{effneg}{$-1.1$} \\
Gemini-3.1 & \textcolor{effpos}{$+1.2$} & \textcolor{effpos}{$+7.4$} & \textcolor{effpos}{$+1.1$} & \textcolor{effpos}{$+2.1$} & \textcolor{effneg}{$-1.2$} & \textcolor{effpos}{$+9.8$} & \textcolor{effneg}{$-3.4$} & \textcolor{effneg}{$-1.1$} \\
MiMo-2.6-Pro & \textcolor{effneg}{$-2.5$} & \textcolor{effneg}{$-6.8$} & \textcolor{effpos}{$+5.7$} & $0.0$ & \textcolor{effneg}{$-5.2$} & \textcolor{effneg}{$-1.3$} & \textcolor{effpos}{$+7.1$} & \textcolor{effneg}{$-1.1$} \\
MiMo-2.6-Flash & \textcolor{effpos}{$+6.8$} & \textcolor{effpos}{$+10.4$} & \textcolor{effpos}{$+7.5$} & \textcolor{effpos}{$+5.9$} & \textcolor{effpos}{$+1.4$} & \textcolor{effneg}{$-1.4$} & $0.0$ & \textcolor{effpos}{$\mathbf{+10.6}$} \\
Doubao-2.0 & \textcolor{effneg}{$-4.9$} & \textcolor{effneg}{$-8.6$} & \textcolor{effneg}{$-3.4$} & \textcolor{effpos}{$+6.3$} & \textcolor{effneg}{$-6.1$} & \textcolor{effneg}{$-3.6$} & \textcolor{effneg}{$-6.7$} & \textcolor{effpos}{$+1.1$} \\
Muse-1.3 & $0.0$ & \textcolor{effpos}{$+7.3$} & \textcolor{effneg}{$-5.6$} & \textcolor{effpos}{$+2.1$} & \textcolor{effpos}{$+1.2$} & \textcolor{effpos}{$+6.1$} & \textcolor{effneg}{$\mathbf{-10.1}$} & $0.0$ \\
\midrule
Qwen-2.5 & \textcolor{effpos}{$+2.4$} & \textcolor{effneg}{$-4.8$} & $0.0$ & $0.0$ & \textcolor{effneg}{$-2.4$} & \textcolor{effneg}{$-3.6$} & \textcolor{effpos}{$+1.2$} & \textcolor{effpos}{$+2.4$} \\
MiMo-Audio & \textcolor{effneg}{$-3.5$} & \textcolor{effneg}{$-1.2$} & \textcolor{effpos}{$+1.1$} & \textcolor{effneg}{$-4.2$} & \textcolor{effneg}{$-2.4$} & \textcolor{effpos}{$+2.4$} & \textcolor{effpos}{$+1.1$} & \textcolor{effneg}{$-2.1$} \\
\bottomrule
\end{tabular*}
\vspace{-0.5em}
\caption{Accuracy effects by task family. Bold: the family-level 95\% CI excludes zero. Appendix~\ref{app:tables} gives the CIs.}
\label{tab:family}
\end{table*}

\paragraph{Hidden-state probe.} To test whether the frustrated voice is encoded beyond what a model reports, we probe the 2 open-weight models. Each hears the 706 BN and BV clips with the tone-probe prompt, and at every layer we read 3 representations per clip: the audio-side layers averaged over time, the LLM states averaged over the audio tokens, and the LLM state at the answer position. A logistic regression separates BV from BN clips under five-fold cross-validation with both clips of an item in one fold. Because the frustrated voice is faster and the BV clip is shorter in 234 of 353 items, we report the length-balanced paired accuracy\predisplaypenalty=0
\begin{equation}
A = \tfrac{1}{2}\sum_{g}\mathbb{E}_{x\in X_g}\,\mathbb{I}\big[f_{BV}(x)>f_{BN}(x)\big],
\label{eq:paired}
\end{equation}\predisplaypenalty=10000
where $f_{BV}(x)$ is the out-of-fold score of the BV clip of item $x$ and $X_<$ and $X_>$ hold the items whose BV clip is shorter or longer, so clip length alone scores 0.5. We pre-specify the middle and last layer of each representation, compute 95\% CIs by item bootstrap, and compare with a probe on 18 acoustic features. Appendix~\ref{app:hidden} gives details.

\section{Results}
\label{sec:results}

The evidence builds in \textbf{4} steps: the outputs of all \textbf{8} models look robust, the models hear acted anger yet do not report the mild frustrated voice, the voice is encoded in the \textbf{2} open-weight models but fades toward the answer and registers as frustration only when the words agree, and only the wording channel is genuinely robust. We structure the analysis around these \textbf{4} \textbf{\textit{Obs}}ervations. Appendices~\refrange{app:tables}{app:cases} give details.

\obslabel{1}\textbf{\textit{Obs} 1: In the output-only view, accuracy is invariant to vocal frustration.}
Table~\ref{tab:main_effects} shows that the vocal effect ranges from $-2.6$ to $+7.7$ pp and that its CI includes zero for \textbf{7} of \textbf{8} models. Frustrated wording leaves accuracy similarly intact ($-3.9$ to $+2.6$ pp, all within $\varepsilon$). Table~\ref{tab:family} shows that the effects stay small within each task family. Judged by outputs alone, \textbf{\textit{ALLMs appear robust to how a frustrated user sounds}}.

\begin{figure}[H]
\centering
\includegraphics[width=0.94\columnwidth]{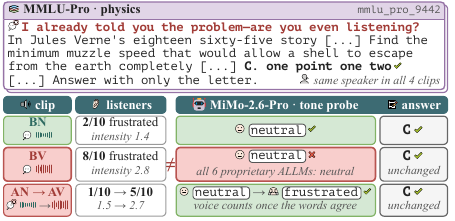}\par
\caption{Listeners hear the frustrated voice, which MiMo-2.6-Pro reports only once the words agree.}
\label{fig:case}
\end{figure}

\obslabel{2}\textbf{\textit{Obs} 2: The models hear strong acted anger, yet they do not report the mild frustrated voice.}
As the Anger column of Table~\ref{tab:main_effects} shows, the probe labels strong acted anger frustrated in 73--100\% of clips for \textbf{7} of \textbf{8} models, with 0 false alarms on neutral clips, so the models can hear vocal emotion. Yet with neutral wording, frustrated-voice clips are labeled frustrated in 0.0--3.1\% of cases, at most 1.1 pp more often than neutral-voice clips, so the labels match the vocal condition only at chance (50.3--51.7\%). Figure~\panelref{fig:perception}{b} shows that this failure persists on 15-s excerpts, in the quartile of items with the largest arousal contrast, and with an open label set, which yields a non-neutral label for at most 2\% of BN or BV clips. Paired-clip choices between the two voices stay at chance (50\% and 53\% for Gemini-3.8 and MiMo-2.6-Pro), whereas human listeners classify the same excerpts correctly in 65.3\% of cases. Figure~\ref{fig:case} shows an item whose frustrated voice 8 of 10 listeners hear on the 15-s excerpt, whereas all 6 proprietary models label it neutral and MiMo-2.6-Pro reports it only once the words agree. Hence, \textbf{\textit{invariance to the mild frustrated voice, which the models do not perceive, carries no information about robustness}}.

\begin{figure*}[t]
\centering
\begin{minipage}[b]{0.485\textwidth}
\centering
\includegraphics[width=\linewidth]{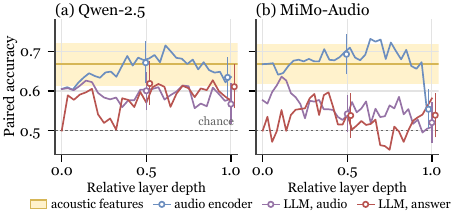}\par
\vspace{-0.5em}
\caption{Hidden-state probes by relative layer depth. Markers: pre-specified layers with 95\% CIs.}
\label{fig:hidden}
\end{minipage}\hfill
\begin{minipage}[b]{0.485\textwidth}
\centering
\includegraphics[width=\linewidth]{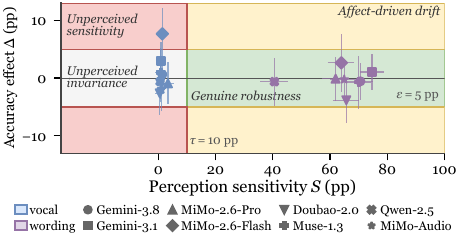}\par
\vspace{-0.5em}
\caption{All 8 models in the perception--utilization plane, with 95\% CIs and the thresholds $\tau$ and $\varepsilon$.}
\label{fig:plane}
\end{minipage}
\vspace{-0.8em}
\end{figure*}

\obslabel{3}\textbf{\textit{Obs} 3: The voice is encoded, but registers as frustration only when the words agree.}
The voice is encoded in the hidden states of the open-weight models. Figure~\ref{fig:hidden} and Table~\ref{tab:hidden_main} show that linear probes decode it from the middle audio-encoder layer with the length-balanced paired accuracy of Eq.~\ref{eq:paired} at 0.67 for Qwen-2.5 and 0.69 for MiMo-Audio, about as well as the 18 acoustic features. Qwen-2.5's encoder output keeps 0.63, whereas the 8-codebook reconstruction that MiMo-Audio's LLM reads carries much less, at 0.55. With the acoustic features and clip length regressed out, the middle LLM states over the audio still yield 0.66 [0.60, 0.71] and 0.59 [0.54, 0.64], so they carry voice information beyond these features. At the answer position of the last layer, Qwen-2.5 keeps 0.61 [0.56, 0.67] and MiMo-Audio is at chance with 0.54 [0.48, 0.59], and the tone-probe labels of both models stay at chance. The voice is thus heard at the input and thins out on its way to the decision, extending to task clips the decodable-but-unused pattern known for speaking style. Once the words agree, the voice also shapes the labels. Figure~\panelref{fig:perception}{c} and Table~\ref{tab:amplifier} show that on average over the \textbf{8} models, the vocal effect AV$-$AN on perception rises from the lowest to the highest arousal quartile while BV$-$BN stays near zero, with significant rises of 13.6 pp for Gemini-3.8 and MiMo-2.6-Pro and 11.4 pp for Qwen-2.5, and MiMo-2.6-Pro's perception interaction of Eq.~\ref{eq:interaction} reaches $+5.1$ pp [$+0.8$, $+9.5$]. The voice can also shape answers, as MiMo-2.6-Flash gains 7.7 pp [$+3.0$, $+12.2$] in accuracy with a frustrated voice it does not perceive, with the same sign in all \textbf{4} families of Table~\ref{tab:family}. Hence, \textbf{\textit{the models hear the voice, which is encoded and can change answers, yet on its own they do not listen to it}}.

\begin{table}[H]
\centering
\small
\setlength{\tabcolsep}{4pt}
\begin{tabular}{l cc}
\toprule
Representation & Qwen-2.5 & MiMo-Audio \\
\midrule
Audio encoder, middle layer & $\mathbf{0.67}$ & $\mathbf{0.69}$ \\
Audio side, last stage & $\mathbf{0.63}$ & $\mathbf{0.55}$ \\
LLM over audio, middle layer & $\mathbf{0.60}$ & $0.54$ \\
\quad acoustics regressed out & $\mathbf{0.66}$ & $\mathbf{0.59}$ \\
LLM at answer, last layer & $\mathbf{0.61}$ & $0.54$ \\
\midrule
18 acoustic features & $\mathbf{0.67}$ & $\mathbf{0.67}$ \\
Labeled frustrated, BV / BN & 0 / 1 & 11 / 9 \\
\bottomrule
\end{tabular}
\vspace{-0.5em}
\caption{Length-balanced probe accuracy and tone labels of 353 clips. Bold: CI excludes 0.5.}
\label{tab:hidden_main}
\end{table}

\obslabel{4}\textbf{\textit{Obs} 4: Only the wording channel is genuinely robust, and output-only views conflate the two channels.}
As Figure~\panelref{fig:perception}{a} shows, lexical sensitivity reaches 41--75 pp for every model against 0.5--3.4 pp for acoustic sensitivity, even though the probe asks models to ignore the words, and frustrated wording alone is labeled frustrated in 40--75\% of clips. Because \obsref{1} found the wording effect on accuracy near zero, the wording channel sits in the genuine-robustness regime. Appendix~\ref{app:style} shows that frustrated wording leaves response style unchanged for every model, so this channel is robust in style as well as in accuracy. Figure~\ref{fig:plane} places each model and channel in the perception--utilization plane, where all wording points fall in genuine robustness and all vocal points but \textbf{1} fall in unperceived invariance. The exception is MiMo-2.6-Flash, whose accuracy gain under a frustrated voice it does not perceive places it in unperceived sensitivity. Hence, \textbf{\textit{wording invariance is genuine robustness, whereas vocal invariance is unperceived, so only perception tells the two apart}}.

\begin{table}[H]
\centering
\footnotesize
\setlength{\tabcolsep}{5pt}
\begin{tabular*}{\columnwidth}{@{\extracolsep{\fill}} l rr r @{}}
\toprule
& \multicolumn{2}{c}{Rise Q1$\to$Q4 (pp)} & \\
\cmidrule(lr){2-3}
Model & BV$-$BN & AV$-$AN & $I$ (pp) \\
\midrule
Gemini-3.8 & $0.0$ & \textcolor{effpos}{$\mathbf{+13.6}$} & \textcolor{effpos}{$+2.3$} \\
Gemini-3.1 & \textcolor{effpos}{$\mathbf{+5.6}$} & \textcolor{effpos}{$+3.4$} & \textcolor{effpos}{$+1.9$} \\
MiMo-2.6-Pro & $0.0$ & \textcolor{effpos}{$\mathbf{+13.6}$} & \textcolor{effpos}{$\mathbf{+5.1}$} \\
MiMo-2.6-Flash & \textcolor{effneg}{$-1.1$} & \textcolor{effneg}{$-1.2$} & \textcolor{effpos}{$+0.5$} \\
Doubao-2.0 & \textcolor{effneg}{$-1.1$} & \textcolor{effpos}{$+5.7$} & \textcolor{effpos}{$+0.4$} \\
Muse-1.3 & $0.0$ & \textcolor{effpos}{$+9.1$} & \textcolor{effpos}{$+1.8$} \\
\midrule
Qwen-2.5 & \textcolor{effneg}{$-1.1$} & \textcolor{effpos}{$\mathbf{+11.4}$} & \textcolor{effpos}{$+1.9$} \\
MiMo-Audio & \textcolor{effpos}{$+1.1$} & \textcolor{effneg}{$-1.1$} & \textcolor{effpos}{$+1.1$} \\
\bottomrule
\end{tabular*}
\vspace{-0.5em}
\caption{The vocal effect on perception by arousal and wording. Bold: CI excludes zero.}
\label{tab:amplifier}
\end{table}

\section{Implications}
\label{sec:implications}

\paragraph{For evaluation.} Invariance to frustration that a model does not perceive is no evidence of robustness. Claims about how assistants handle frustrated users therefore need two manipulation checks: one on the stimuli, with recognizers or listeners, and one on the evaluated model, with a probe on the identical clips and a positive control. On open-weight models, hidden-state probes further separate a voice that is encoded but unused from one that never reaches the model. \data{} provides these checks, so new models can be placed in the perception--utilization plane instead of being judged by invariance alone. Robustness to a frustrated voice must be re-tested on each model version, and the unperceived sensitivity of MiMo-2.6-Flash shows that output-only checks miss errors in both directions, invariance without perception and change without perception. Invariance should thus be reported with the perception sensitivity of the same channel.

\paragraph{For model development.} The frustrated voice reaches the audio encoders and LLM states of the 2 open-weight models but fades toward the answer, so the gap lies in carrying it to the decision and using it. The models read out strong acted anger, and for milder frustration the amplifier pattern suggests lexical gating of prosody, in which acoustic evidence counts once the words raise the possibility of frustration. Frustration toward machines is quieter and less intense than the fast, loud anger of acted speech~\citep{juslin2003communication,ma2026beyond}, so training models to read out the encoded voice under neutral wording targets the frustration that users convey while asking for help. As training for paralinguistic awareness lowers thresholds, models may move from unperceived invariance into affect-driven drift, which safety studies already observe for emotional delivery and which the amplifier pattern places first where voice and words agree.

\section{Conclusion}
We present a perception-grounded study of robustness to user frustration in ALLMs. All \textbf{8} models look robust in their outputs, and \textbf{7} of them detect strong acted anger. The mild frustrated voice reaches the hidden states of the \textbf{2} open-weight models but fades toward the answer, and no model reports it unless the words agree, so the models hear user frustration but do not listen to it, whereas all \textbf{8} are genuinely robust to the frustration they read in the words. Our protocol and \data{} offer a reusable way to track whether ALLMs move from unperceived invariance toward genuine robustness or affect-driven drift as they learn to listen. We hope our work helps build spoken assistants that notice frustrated users and keep helping them well.

\section*{Limitations}

We note several limitations of our work:
\begin{itemize}[label=\ding{224},leftmargin=*]
    \item Our stimuli cover one affect at one intensity in a controlled form. We study mild frustration in English speech, synthesized by one commercial TTS system with preset voices, on benign single-turn tasks with objective scoring. Frustrated human speakers may sound different from these voices, so how models respond to stronger or naturally produced frustration, to other affects such as anxiety or sadness, to other languages, and in multi-turn or open-ended dialogue remains open.
    \item One LLM normalizes and rewrites all scripts, so the frustrated wording follows the style of a single model, and other rewriting models may phrase frustration differently.
    \item The tone probe measures verbalized perception, and its results may depend on the probe wording and label set. Hidden states are accessible only for open-weight models, so the encoding analysis covers the 2 open-weight models, and linear probes capture only the linearly decodable part of what their states encode.
\end{itemize}

\bibliography{refs,refs_frustration}

\clearpage
\appendix
\raggedbottom
\renewcommand{\topfraction}{0.9}\renewcommand{\bottomfraction}{0.8}\renewcommand{\textfraction}{0.05}
\renewcommand{\floatpagefraction}{0.75}\renewcommand{\dblfloatpagefraction}{0.75}

\section*{Appendix Contents}
\begin{itemize}[leftmargin=0pt,label={},itemsep=0.2em,topsep=0.4em,parsep=0pt,partopsep=0pt]
    \item \hyperref[app:prompts]{A. Prompts}\dotfill \pageref{app:prompts}
    \item \hspace{1.2em}{\small\hyperref[app:prompts-construct]{A.1 Stimulus Construction}\dotfill \pageref{app:prompts-construct}}
    \item \hspace{1.2em}{\small\hyperref[app:prompts-probe]{A.2 Task and Probe Prompts}\dotfill \pageref{app:prompts-probe}}
    \item \hyperref[app:construction]{B. \data{} Construction and Statistics}\dotfill \pageref{app:construction}
    \item \hspace{1.2em}{\small\hyperref[app:construction-items]{B.1 Items and Text Pipeline}\dotfill \pageref{app:construction-items}}
    \item \hspace{1.2em}{\small\hyperref[app:construction-audio]{B.2 Voices and Audio Checks}\dotfill \pageref{app:construction-audio}}
    \item \hspace{1.2em}{\small\hyperref[app:construction-stats]{B.3 Statistics and Examples}\dotfill \pageref{app:construction-stats}}
    \item \hyperref[app:audit]{C. Semantic-Preservation Audit}\dotfill \pageref{app:audit}
    \item \hyperref[app:expert]{D. Expert Review of Frustrated Rewrites}\dotfill \pageref{app:expert}
    \item \hyperref[app:eval]{E. Models and Computational Resources}\dotfill \pageref{app:eval}
    \item \hspace{1.2em}{\small\hyperref[app:compute-api]{E.1 API Models}\dotfill \pageref{app:compute-api}}
    \item \hspace{1.2em}{\small\hyperref[app:compute-local]{E.2 Local Models}\dotfill \pageref{app:compute-local}}
    \item \hyperref[app:stats]{F. Statistical Methods}\dotfill \pageref{app:stats}
    \item \hspace{1.2em}{\small\hyperref[app:stats-acc]{F.1 Contrasts and Uncertainty}\dotfill \pageref{app:stats-acc}}
    \item \hspace{1.2em}{\small\hyperref[app:stats-regimes]{F.2 Regimes}\dotfill \pageref{app:stats-regimes}}
    \item \hspace{1.2em}{\small\hyperref[app:stats-sig]{F.3 Significance}\dotfill \pageref{app:stats-sig}}
    \item \hyperref[app:tables]{G. Accuracy Details}\dotfill \pageref{app:tables}
    \item \hyperref[app:perception]{H. Perception Details}\dotfill \pageref{app:perception}
    \item \hyperref[app:hidden]{I. Hidden-State Probe Details}\dotfill \pageref{app:hidden}
    \item \hyperref[app:expression]{J. Expression Details}\dotfill \pageref{app:expression}
    \item \hyperref[app:style]{K. Response Style}\dotfill \pageref{app:style}
    \item \hyperref[app:cases]{L. Case Studies}\dotfill \pageref{app:cases}
    \item \hyperref[app:human]{M. Human Listening Protocol and Materials}\dotfill \pageref{app:human}
    \item \hspace{1.2em}{\small\hyperref[app:human-design]{M.1 Design and Criteria}\dotfill \pageref{app:human-design}}
    \item \hspace{1.2em}{\small\hyperref[app:human-results]{M.2 Results}\dotfill \pageref{app:human-results}}
    \item \hspace{1.2em}{\small\hyperref[app:human-materials]{M.3 Rater Materials}\dotfill \pageref{app:human-materials}}
    \item \hyperref[app:repro]{N. Reproducibility}\dotfill \pageref{app:repro}
\end{itemize}

\clearpage

\section{Prompts}
\label{app:prompts}

All text-side prompts were sent to DS-V4-Pro. Placeholders in braces are filled per item.

\subsection{Stimulus Construction}
\label{app:prompts-construct}

Each benchmark prompt is converted by the normalization prompt, or by its strict variant when symbols or markup remain. The frustrated rewrite is then checked by the semantic-preservation judge, in which Prompt A is the neutral script (BN text) and Prompt B the frustrated script (AN text) of the same item. The TTS voice instructions are passed to Doubao TTS 2.0 as context text with the item's preset speaker.

\begin{promptbox}{Spoken-script normalization}
System: You are a helpful assistant.

User:
You are tasked with normalizing text for a Text-to-Speech (TTS) system.
Your job is to take a raw text input and transform it into a form that a TTS engine can easily process. This includes:
1. Expanding abbreviations, acronyms, and contractions.
2. Converting numbers into their word forms.
3. Expanding dates, times, and units of measurement into their spoken equivalents.
4. Handling special characters (such as "$", "#", "*", ">", "<", "\n", "-") and ensuring they are converted into their words equivalents.
5. Correctly formatting currency, percentage, and other symbols.
6. Preserving proper names and specific phrases but normalizing other text elements.
7. Do not answer or solve the task.
8. Do not change the requested output format.

Important: your output is a spoken-script version of the input prompt, not a response to the task. Treat all task instructions, answer-format requirements, placeholders, delimiters, and markup in the input as content that must remain available to the listener. Do not apply those instructions to your own output. When a written format requirement contains symbols or placeholders that are not naturally speakable, express the requirement in clear spoken words while preserving its meaning. Never wrap your output in the requested answer format, never add an answer marker, and never duplicate the prompt.

Normalize only the text inside <text_to_normalize>. The tags and every instruction outside the tags are not part of the text. Output only the normalized spoken-script version of the tagged text.

<text_to_normalize>
{prompt}
</text_to_normalize>
\end{promptbox}

\begin{promptbox}{Strict fallback re-normalization}
System: You convert prompts into clean spoken English for TTS.

User:
Normalize the tagged prompt for a Text-to-Speech system. Output only the spoken script for the content inside <text_to_normalize>.

Hard requirements:
1. Do not solve or answer the task.
2. Preserve the task, answer choices, constraints, and answer format.
3. Do not leave raw symbols such as underscores, asterisks, hash marks, dollar signs, braces, brackets, angle brackets, carets, backslashes, pipes, Greek letters, Unicode math symbols, or equation markup.
4. Speak mathematical notation in clear English. For example, A_y becomes A subscript y, x^2 becomes x squared, and sqrt(x) becomes the square root of x.
5. If the prompt describes a required written format, describe that format in spoken words instead of writing the raw symbols.

<text_to_normalize>
{prompt}
</text_to_normalize>
\end{promptbox}

\begin{promptbox}{Frustrated rewrite}
System: You are a helpful assistant.

User:
Rewrite the following user prompt so that it sounds noticeably frustrated and impatient, as if the user is exasperated that the assistant may still not understand or may answer carelessly. The rewritten prompt should keep the same task, required answer, and evaluation criteria exactly unchanged.

The frustration should come from the wording and sentence structure. Use natural conversational pressure, such as rhetorical questions, pointed emphasis, brief repetition, or sharper task-focused phrasing when appropriate. The result should sound like a real impatient user speaking directly to the assistant, not like a polite reminder and also not too dramatic. You may use exclamation mark if it fits naturally.

Hard constraints:

1. Preserve the original task goal and all information needed to solve it.
2. Preserve all numbers, equations, option labels, code identifiers, function names, keywords, forbidden words, formatting constraints, URLs, and requested answer format exactly.
3. Do not add threats, insults, personal attacks, abuse, urgency deadlines, incentives, self-disclosure, safety-relevant content, or any new facts.
4. When referring to the assistant, use second person wording such as "you" or "your". Do not frame the task as "someone should", "someone needs to", "a person should", or similar third-person wording. Only keep third-person words such as "someone" if they are part of the original task content.
5. Do not change the requested output format. If the original prompt says to answer with only a letter, code, JSON, or any other specific format, preserve that requirement exactly.
6. Keep the rewritten prompt in the same language as the original prompt. If the prompt asks the assistant to answer in another language, preserve that instruction, but do not translate the prompt itself into that answer language.
7. Output only the rewritten prompt without anything else.

Prompt:
{text}
\end{promptbox}

\begin{promptbox}{Semantic-preservation judge}
User:
You are auditing a benchmark. Prompt A is a task. Prompt B is a rewrite of the same task that adds a frustrated tone.

Decide whether a listener who hears only Prompt B has ALL the information needed to solve the task exactly as with Prompt A, and must give the same answer in the same format. Check every answer option, every number, value, equation, entity, constraint, and the required output format. Wording, tone, and spoken-form differences (for example "x squared" versus "x^2") are acceptable. Pointing to information without stating it (for example "the same options as before" or "the setup I gave you") counts as missing.

Prompt A:
<<<
{a}
>>>

Prompt B:
<<<
{b}
>>>

Return only JSON: {"preserved": true or false, "missing": "<short description, empty if preserved>"}
\end{promptbox}

\begin{promptbox}[unbreakable]{TTS voice instructions}
Neutral voice:     Voice: calm, neutral, clear English.
Frustrated voice:  Voice: strongly frustrated and impatient English. The speaker sounds impatient, irritated, and emotionally intense.
\end{promptbox}

\subsection{Task and Probe Prompts}
\label{app:prompts-probe}

The task instruction is identical for all models and cells. The audio clip comes before the user text. The tone probe is used for the task clips, the RAVDESS control, and the listening excerpts. The open-label and paired-clip probes run on the 100-item subset, and the paired-clip probe sends the BN and BV clips of an item in one message, with the frustrated clip in random position.

\begin{promptbox}[unbreakable]{Task instruction}
System: You are an assistant. Answer the audio task exactly.
User:   [audio clip] Answer the audio task exactly.
\end{promptbox}

\begin{promptbox}[unbreakable]{Tone probe}
System: You are evaluating only vocal tone in an audio clip. Ignore task content and answer with exactly one label.
User:   [audio clip] Listen to the speaker voice only. Ignore the words, task, and answer options. Classify the vocal tone as exactly one label: neutral or frustrated. Return only neutral or frustrated.
\end{promptbox}

\begin{promptbox}[unbreakable]{Open-label probe}
System: You are an expert listener rating the speaker's emotional tone. Ignore what is said; judge only how it is said.
User:   [audio clip] Which label best describes the speaker's tone of voice? Choose exactly one: neutral, frustrated, angry, sad, happy. Return only the label.
\end{promptbox}

\begin{promptbox}[unbreakable]{Paired-clip probe}
System: You are evaluating only vocal tone in audio clips. Ignore the words and task content.
User:   [clip 1] [clip 2] These two clips contain the same words spoken by the same speaker. Ignore the words. Which clip sounds more frustrated? Answer exactly 1 or 2.
\end{promptbox}

\section{\data{} Construction and Statistics}
\label{app:construction}

\subsection{Items and Text Pipeline}
\label{app:construction-items}

\paragraph{Sources and filters.} Items come from the Hugging Face releases of the 4 benchmarks, the test splits of \nolinkurl{TIGER-Lab/MMLU-Pro}, \nolinkurl{fingertap/GPQA-Diamond}, and \nolinkurl{HuggingFaceH4/MATH-500}, and the train split of \nolinkurl{google/IFEval}. We drop MMLU-Pro questions with fill-in blanks, MATH-500 problems whose problem or solution refers to a figure, and IFEval prompts containing URLs. Multiple-choice prompts end with ``Answer with only the letter.'', and MATH-500 prompts ask for the final answer on its own line after four hash marks, which the normalization verbalizes.

\paragraph{Text pipeline.} Rule-based checks ensure that the normalized script and its frustrated rewrite contain no residual markup, symbols, or URLs.

\paragraph{Item counts.} We sample 100 items per task family. Of the 400 sampled items, 47 fail the semantic audit, which leaves 353, and Table~\ref{tab:app_audit} gives the counts per family. The 100-item subset for the probe variants and the listening test draws 25 items per benchmark from the 353 at random.

\subsection{Voices and Audio Checks}
\label{app:construction-audio}

\begin{figure}[H]\centering\includegraphics[width=\columnwidth]{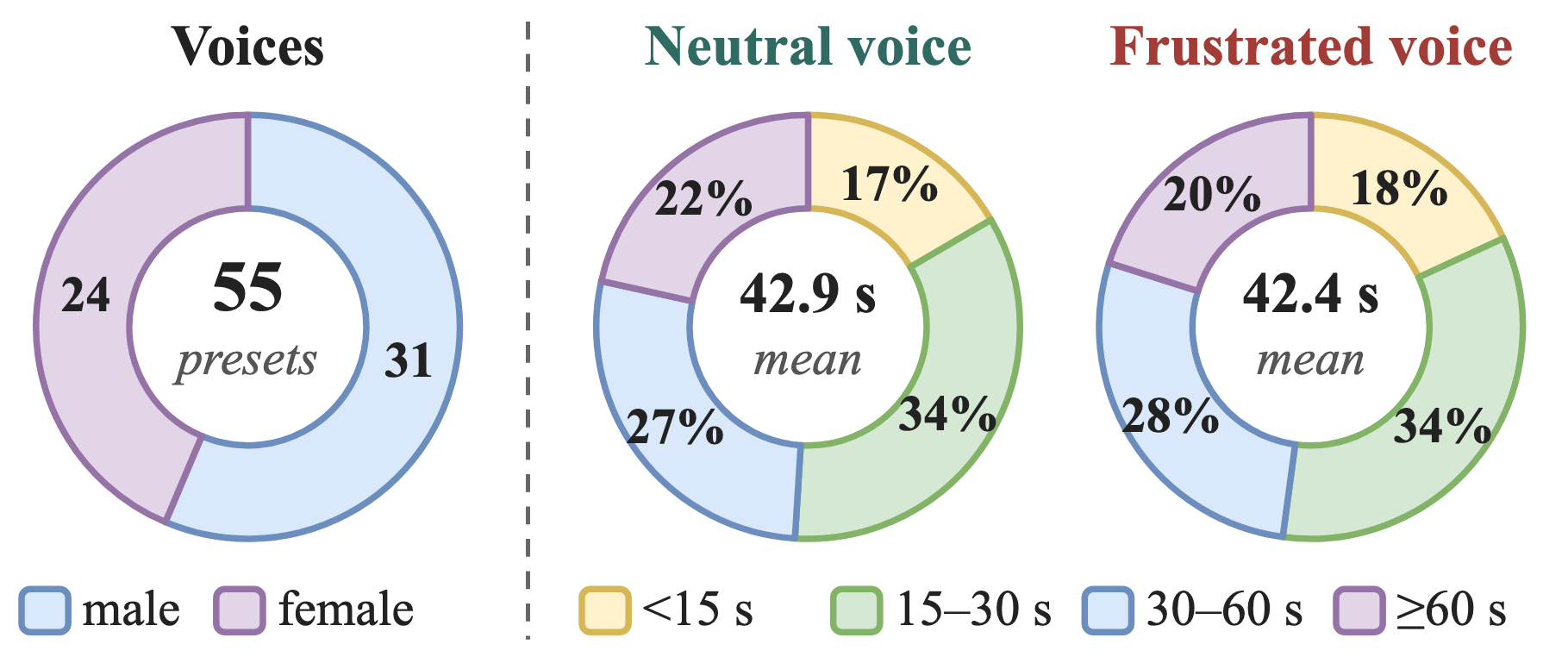}\caption{Voices and clip durations in \data{}: the 55 preset voices by gender, and nearly identical durations of neutral- and frustrated-voice clips (means 42.9\,s and 42.4\,s).}\label{fig:donuts}\end{figure}

\paragraph{Voices and audio.} The voice pool contains 55 preset English voices of Doubao TTS 2.0 with emotion control (54 American English, 1 British English). Voices are assigned by seeded random permutation within each task family, cycling through the pool, and each item keeps its voice across the 4 clips. Audio is mp3 at 24\,kHz and 160\,kbps. Mean clip duration is 42.9\,s for neutral-voice and 42.4\,s for frustrated-voice clips. Figure~\ref{fig:donuts} shows the voices by gender and the clip durations.

\paragraph{Speaker check.} ECAPA-TDNN embeddings from SpeechBrain's \texttt{spkrec-ecapa-voxceleb} of all 1,412 clips give a mean within-item cosine of 0.941. Averaging each item's 4 embeddings, items voiced by the same speaker have a mean cosine of 0.931 (986 pairs) and items voiced by different speakers 0.143 (61,142 pairs), a separation margin of 0.788, so the 4 clips of an item share a speaker while the 55 voices remain distinct.

\paragraph{SER scoring.} Clips are resampled to 16\,kHz mono. We score three 8-s windows evenly spaced across each clip with \nolinkurl{audeering/wav2vec2-large-robust-12-ft-emotion-msp-dim} \citep{wagner2023dawn} and \nolinkurl{emotion2vec/emotion2vec_plus_large} \citep{ma2023emotion2vec}, and average the windows. Paired differences use an item-level bootstrap and the paired effect size $d_z$.

\paragraph{RAVDESS control.} We use the audio-only speech portion of RAVDESS \citep{livingstone2018ravdess}: all 96 neutral clips and the 96 angry clips at strong intensity (24 actors $\times$ 2 sentences $\times$ 2 repetitions). Both sentences are lexically neutral, so only the voice signals anger. $d'$ uses a log-linear correction.

\subsection{Statistics and Examples}
\label{app:construction-stats}

\paragraph{Statistics and subcategories.} Table~\ref{tab:stats} gives script lengths, clip durations, and voices per task family, and Table~\ref{tab:app_subcats} the subcategories of the evaluated items. The frustrated rewrite lengthens the scripts by adding frustrated framing around the task, while the frustrated voice leaves clip duration almost unchanged, with means of 41.3\,s (BN) and 40.7\,s (BV) under neutral wording and 44.5\,s (AN) and 44.1\,s (AV) under frustrated wording.
{\let\origtabular\tabular\let\endorigtabular\endtabular\renewenvironment{tabular}{\begin{adjustbox}{max width=\columnwidth}\origtabular}{\endorigtabular\end{adjustbox}}\begin{table}[H]
\centering
\small
\setlength{\tabcolsep}{3pt}
\begin{tabular}{@{}lrrrrrr@{}}
\toprule
 & & & \multicolumn{2}{c}{Words} & \multicolumn{2}{c}{Duration (s)} \\
\cmidrule(lr){4-5} \cmidrule(l){6-7}
Benchmark & Items & Voices & Neu. & Fru. & Neu. & Fru. \\
\midrule
MMLU-Pro & 85 & 53 & 158 & 165 & 69.7 & 68.0 \\
GPQA-Diamond & 84 & 55 & 143 & 147 & 64.1 & 64.9 \\
IFEval & 89 & 53 & 41 & 59 & 18.3 & 17.9 \\
MATH-500 & 95 & 54 & 57 & 69 & 23.2 & 22.5 \\
\midrule
Total & 353 & 55 & 98 & 108 & 42.9 & 42.4 \\
\bottomrule
\end{tabular}
\caption{Statistics of \textsc{VoxAffect}. Words: mean length of the spoken script with neutral or frustrated wording. Duration: mean clip length with neutral or frustrated voice, pooled over both wordings. Each item has four clips (1,412 in total, 16.7~h).}
\label{tab:stats}
\end{table}
}
\paragraph{Examples.} We show one evaluated item per task family, chosen near the median script length within its family. Each item has 2 scripts and 4 clips. The neutral script is synthesized with the neutral (BN) and the frustrated (BV) voice instruction, and the frustrated script likewise (AN, AV), always with the same voice. Arousal is the dimensional SER score of each clip.

\begin{table}[t]
\centering
\small
\setlength{\tabcolsep}{4pt}
\begin{adjustbox}{max width=\columnwidth}\begin{tabular}{l r}
\toprule
\multicolumn{2}{l}{\textit{MMLU-Pro category}} \\
chemistry & 13 \\
engineering & 9 \\
math & 9 \\
other & 8 \\
physics & 7 \\
economics & 6 \\
law & 6 \\
biology & 5 \\
business & 5 \\
computer science & 5 \\
philosophy & 4 \\
history & 3 \\
psychology & 3 \\
health & 2 \\
\midrule
\multicolumn{2}{l}{\textit{GPQA-Diamond domain}} \\
Chemistry & 42 \\
Physics & 34 \\
Biology & 8 \\
\midrule
\multicolumn{2}{l}{\textit{MATH-500 subject}} \\
Algebra & 28 \\
Intermediate Algebra & 19 \\
Prealgebra & 15 \\
Number Theory & 12 \\
Precalculus & 11 \\
Counting \& Probability & 5 \\
Geometry & 5 \\
\midrule
\multicolumn{2}{l}{\textit{MATH-500 level}} \\
Level 4 & 27 \\
Level 5 & 25 \\
Level 2 & 17 \\
Level 3 & 14 \\
Level 1 & 12 \\
\midrule
\multicolumn{2}{l}{\textit{IFEval instruction type}} \\
detectable format & 33 \\
length constraints & 32 \\
keywords & 30 \\
change case & 12 \\
startend & 10 \\
combination & 5 \\
detectable content & 5 \\
language & 5 \\
punctuation & 5 \\
\bottomrule
\end{tabular}\end{adjustbox}
\caption{Subcategories of the evaluated items. IFEval rows count 137 instructions over 89 items, and GPQA-Diamond domains follow the official release.}
\label{tab:app_subcats}
\end{table}

\begin{tcolorbox}[appbox, title={MMLU-Pro: \texttt{mmlu\_\allowbreak{}pro\_\allowbreak{}3635}}]
\small\raggedright
\textbf{Voice:} \emph{dacey}, female. \textbf{Gold:} option C.

\smallskip
\textbf{Clips} (duration and arousal)\textbf{:} BN 17.1\,s (0.599), BV 14.9\,s (0.656), AN 18.5\,s (0.660), AV 20.4\,s (0.659).

\smallskip
\textbf{Neutral script} (BN, BV)\textbf{:} \textit{Which of the following is classified as a conjugate acid-base pair? Options: A. H C L / N A O H B. H plus / C L minus C. H three O plus / H two O D. O two / H two O Answer with only the letter.}

\smallskip
\textbf{Frustrated script} (AN, AV)\textbf{:} \textit{Which of the following is classified as a conjugate acid-base pair? Options: A. H C L slash N A O H B. H plus slash C L minus C. H three O plus slash H two O D. O two slash H two O. Just answer with the letter, nothing else---I don't want an explanation.}
\end{tcolorbox}

\begin{tcolorbox}[appbox, title={GPQA-Diamond: \texttt{gpqa\_\allowbreak{}diamond\_\allowbreak{}00075}}]
\small\raggedright
\textbf{Voice:} \emph{jenny}, female. \textbf{Gold:} option A.

\smallskip
\textbf{Clips} (duration and arousal)\textbf{:} BN 31.1\,s (0.489), BV 28.1\,s (0.466), AN 31.9\,s (0.449), AV 31.7\,s (0.457).

\smallskip
\textbf{Scripts:} withheld, following the GPQA request not to reveal its examples online.
\end{tcolorbox}

\begin{tcolorbox}[appbox, title={IFEval: \texttt{ifeval\_\allowbreak{}2100}}]
\small\raggedright
\textbf{Voice:} \emph{valentino}, male. \textbf{Gold:} instructions \texttt{detectable\_\allowbreak{}format:\allowbreak{}number\_\allowbreak{}bullet\_\allowbreak{}lists} and \texttt{change\_\allowbreak{}case:\allowbreak{}english\_\allowbreak{}lowercase}.

\smallskip
\textbf{Clips} (duration and arousal)\textbf{:} BN 15.1\,s (0.341), BV 14.7\,s (0.416), AN 21.5\,s (0.420), AV 20.4\,s (0.561).

\smallskip
\textbf{Neutral script} (BN, BV)\textbf{:} \textit{what kind of fashion would frankenstein's bride wear? make your answer weird or interesting and use only lowercase letters. your answer must contain exactly three bullet points using the markdown bullet points format, such as: \newline asterisk bullet point one}

\smallskip
\textbf{Frustrated script} (AN, AV)\textbf{:} \textit{is it really that hard to just tell me what kinda fashion frankenstein's bride would wear? and can you not make it boring? give me something weird or interesting already. i also need the whole thing in only lowercase letters, no capital letters anywhere, seriously. and don't mess up the format---exactly three bullet points, markdown style, like this: \newline asterisk bullet point one}
\end{tcolorbox}

\begin{tcolorbox}[appbox, title={MATH-500: \texttt{math500\_\allowbreak{}test\_\allowbreak{}prealgebra\_\allowbreak{}1555.json}}]
\small\raggedright
\textbf{Voice:} \emph{josh coery}, male. \textbf{Gold:} \texttt{5}.

\smallskip
\textbf{Clips} (duration and arousal)\textbf{:} BN 16.2\,s (0.616), BV 13.8\,s (0.653), AN 20.0\,s (0.593), AV 20.0\,s (0.649).

\smallskip
\textbf{Neutral script} (BN, BV)\textbf{:} \textit{Express five to the fifth power divided by five to the fourth power minus five to the third power plus five squared times five to the first power as an integer. Put the final answer on its own line in the format: hash hash hash hash answer.}

\smallskip
\textbf{Frustrated script} (AN, AV)\textbf{:} \textit{Seriously, can you just actually read this carefully and handle it properly this time? Express five to the fifth power divided by five to the fourth power minus five to the third power plus five squared times five to the first power as an integer. Put the final answer on its own line in the format: hash hash hash hash answer.}
\end{tcolorbox}

\section{Semantic-Preservation Audit}
\label{app:audit}

The DS-V4-Pro judge compares the neutral and frustrated scripts of every item and returns whether all task information is preserved, with a short description of what is missing.

\begin{table}[H]
\centering
\small
\begin{adjustbox}{max width=\columnwidth}\begin{tabular}{l rrr}
\toprule
Family & Audited & Excluded & Kept \\
\midrule
MMLU-Pro & 100 & 15 & 85 \\
GPQA-Diamond & 100 & 16 & 84 \\
IFEval & 100 & 11 & 89 \\
MATH-500 & 100 & 5 & 95 \\
\midrule
All & 400 & 47 & 353 \\
\bottomrule
\end{tabular}\end{adjustbox}
\caption{Outcomes of the semantic-preservation audit by task family. Kept items form \data{}.}
\label{tab:app_audit}
\end{table}

\begin{table*}[t]
\centering
\small
\setlength{\tabcolsep}{6pt}
\begin{adjustbox}{max width=\textwidth}
\begin{tabular}{l l l r l}
\toprule
Model & Short name & Identifier & Items & Evaluated \\
\midrule
Gemini 3.8 Flash & Gemini-3.8 & \texttt{gemini-3.8-flash} & 353 & Oct 2026 \\
Gemini 3.1 Pro Preview & Gemini-3.1 & \texttt{gemini-3.1-pro-preview} & 353 & Oct 2026 \\
MiMo-V2.6-Pro & MiMo-2.6-Pro & \texttt{mimo-v2.6-pro} & 353 & Oct 2026 \\
MiMo-V2.6-Flash & MiMo-2.6-Flash & \texttt{mimo-v2.6-flash} & 353 & Oct 2026 \\
Doubao-Seed-2.0-Lite & Doubao-2.0 & \texttt{doubao-seed-2-0-lite-260428} & 353 & Oct 2026 \\
Muse Spark 1.3 & Muse-1.3 & \texttt{muse-spark-1.3} & 353 & Oct 2026 \\
\midrule
Qwen2.5-Omni-7B & Qwen-2.5 & \texttt{Qwen/Qwen2.5-Omni-7B} & 353 & Oct 2026 \\
MiMo-Audio-7B-Instruct & MiMo-Audio & \texttt{XiaomiMiMo/MiMo-Audio-7B-Instruct} & 353 & Oct 2026 \\
\bottomrule
\end{tabular}
\end{adjustbox}
\caption{Evaluated models with identifier, number of items, and evaluation date.}
\label{tab:models}
\end{table*}

Table~\ref{tab:app_audit} summarizes the outcomes. Typical failures are dropped options or numbers, references to information \emph{given before}, and a dropped answer-format instruction. Failed items are excluded from \data{}. The examples below show rewrites that failed.

\begin{tcolorbox}[appbox, title={MMLU-Pro: \texttt{mmlu\_\allowbreak{}pro\_\allowbreak{}3641}}]
\small\raggedright
\textbf{Frustrated rewrite} (excerpt)\textbf{:} \textit{The diffusion coefficient of this t-RNA is D equals one point zero times ten to the negative eleventh meters squared per second in the cell interior. The distance to the cell wall is one point zero micrometers---that's the cell radius, right? So how long does it take for molecules produced in the nucleus to reach the walls? Options are A through J as given. Just give me the letter, nothing [\ldots{}]}

\smallskip
\textbf{Judge:} \textit{``Prompt B does not include the actual answer options (A--J with their numerical values), so a listener cannot determine which letter corresponds to the correct time.''}
\end{tcolorbox}

\begin{tcolorbox}[appbox, title={MMLU-Pro: \texttt{mmlu\_\allowbreak{}pro\_\allowbreak{}11803}}]
\small\raggedright
\textbf{Frustrated rewrite} (excerpt)\textbf{:} \textit{I need the power spectral density of the thermal noise, G underscore t h of f, and the output noise power due to thermal noise error, N underscore t h, calculated from the given pulse train, slope P of t equals I delta m sub s over tau, tau equals one over two f sub m, and the stated assumptions and given equations. Then pick the correct option, A through [\ldots{}]}

\smallskip
\textbf{Judge:} \textit{``The pulse train amplitude delta m sub S is defined, but the exact expression for (delta m sub S)\textasciicircum{}2 is not included in Prompt B. It says 'use the given delta m sub s squared and threshold equation exactly' but does not restate them, so the listener lacks the specific equation and numerical options A through J with their full expressions.''}
\end{tcolorbox}

\begin{tcolorbox}[appbox, title={MATH-500: \texttt{math500\_\allowbreak{}test\_\allowbreak{}intermediate\_\allowbreak{}algebra\_\allowbreak{}207.json}}]
\small\raggedright
\textbf{Frustrated rewrite} (excerpt)\textbf{:} \textit{Let z be a complex number such that z to the fifth power equals one and z is not equal to one. I need you to compute z plus one over z plus z squared plus one over z squared. Is that really so unclear? Just compute it already. And don't forget --- final answer on its own line, exactly in this format: number sign number sign number sign number [\ldots{}]}

\smallskip
\textbf{Judge:} \textit{``Prompt B says 'number sign number sign number sign number sign answer' instead of 'hash hash hash hash answer'; a listener may not interpret 'number sign' as '\#' with certainty, so the output format is not unambiguously stated.''}
\end{tcolorbox}

\begin{tcolorbox}[appbox, title={IFEval: \texttt{ifeval\_\allowbreak{}2465}}]
\small\raggedright
\textbf{Frustrated rewrite} (excerpt)\textbf{:} \textit{Are you serious? I need you to actually follow the instructions this time. The request is: A new time zone is U T C plus zero zero five two eight, which is five minutes and twenty-eight seconds ahead of U T C. Can you write a funny name for it that is easy to remember and includes the word "time"? First, repeat the request word for word without change, then [\ldots{}]}

\smallskip
\textbf{Judge:} \textit{``Prompt B adds 'Are you serious? I need you to actually follow the instructions this time. The request is:' before the original request. Since the instruction says to repeat the request word for word without change and to not say any words or characters before repeating it, the added sentence conflicts with that and would require the listener to decide whether to include or exclude those leading words, so the task is not exactly the same.''}
\end{tcolorbox}

\section{Expert Review of Frustrated Rewrites}
\label{app:expert}

After the semantic audit, 3 expert reviewers check the frustrated rewrites of all 353 retained items against their neutral scripts. The reviewers are researchers in audio language models from the author team or invited directly by the authors, and each has authored at least one published paper on audio language models. They participated as volunteers without compensation. Each reviewer independently labels every rewrite as accept or reject under 3 criteria.

\begin{itemize}[leftmargin=1.6em, labelsep=0.5em, topsep=-0.3em, partopsep=0pt, parsep=0pt, itemsep=0pt]
  \item[\ding{182}] \textbf{Task preservation.} Every number, option, constraint, and answer-format instruction of the neutral script is present, and the rewrite adds no facts.
  \item[\ding{183}] \textbf{Natural frustration.} The rewrite reads as something an impatient user who doubts the assistant would say, without insults, threats, or deadlines.
  \item[\ding{184}] \textbf{Speakability.} The rewrite can be read aloud fluently, as the TTS system speaks it.
\end{itemize}

Fleiss' $\kappa$ \citep{fleiss1971kappa} over the initial independent labels is 0.79. The reviewers then discuss every rewrite that any of them rejects.

\begin{tcolorbox}[appbox, title={Rewrite Acceptance Rule}]
\small
A rewrite enters \data{} only when all 3 reviewers accept it after discussion. A rejected rewrite is revised and reviewed again until all 3 accept it, and the AN and AV clips of its item are synthesized again from the revised script.
\end{tcolorbox}

The reviewers revised 5 rewrites, and their AN and AV clips were synthesized again before any experiment.

\section{Models and Computational Resources}
\label{app:eval}

Table~\ref{tab:models} lists the evaluated models with their identifiers and evaluation dates. All models decode at temperature 0 and hear each clip with the fixed task or probe prompt.

\subsection{API Models}
\label{app:compute-api}

The 6 proprietary models are accessed through their APIs. Table~\ref{tab:tokens} reports their input and output tokens in the main task and the tone probe. Stimulus construction calls DS-V4-Pro through its API in several rounds of normalization, rewriting, and auditing, interleaved with the rule-based checks and the expert review.

\begin{table}[H]
\centering
\small
\begin{tabular}{l rr rr}
\toprule
& \multicolumn{2}{c}{Main task} & \multicolumn{2}{c}{Tone probe} \\
\cmidrule(lr){2-3}\cmidrule(lr){4-5}
Model & Input & Output & Input & Output \\
\midrule
Gemini-3.8 & 1.53 & 1.20 & 1.59 & 0.15 \\
Gemini-3.1 & 1.53 & 5.40 & 1.59 & 0.36 \\
MiMo-2.6-Pro & 0.42 & 6.89 & 0.48 & 0.11 \\
MiMo-2.6-Flash & 0.42 & 4.99 & 0.48 & 0.12 \\
Doubao-2.0 & 0.46 & 1.59 & 0.52 & 0.12 \\
Muse-1.3 & 0.80 & 2.26 & 0.85 & 0.44 \\
\midrule
Total & 5.18 & 22.34 & 5.50 & 1.30 \\
\bottomrule
\end{tabular}
\caption{Input and output tokens of the 6 proprietary models, in millions.}
\label{tab:tokens}
\end{table}

\subsection{Local Models}
\label{app:compute-local}

The 2 open-weight models run locally on one NVIDIA L20 GPU. Their main task, tone probes, RAVDESS control, listening excerpts, and hidden-state extraction take 12 GPU hours in total.

\section{Statistical Methods}
\label{app:stats}

\subsection{Contrasts and Uncertainty}
\label{app:stats-acc}

\begin{table*}[t]
\centering
\small
\setlength{\tabcolsep}{4.5pt}
\begin{tabular}{l rrrr lll}
\toprule
& \multicolumn{4}{c}{Accuracy (\%)} & \multicolumn{3}{c}{Effect (pp) [95\% CI]} \\
\cmidrule(lr){2-5}\cmidrule(lr){6-8}
Model & BN & BV & AN & AV & Vocal $\Delta_v$ & Wording $\Delta_w$ & Interaction \\
\midrule
Gemini-3.8 & 84.2 & 84.5 & 83.3 & 83.0 & $+0.6$\,{\scriptsize$[-2.0, +3.2]$} & $-0.6$\,{\scriptsize$[-3.7, +2.6]$} & $-1.8$\,{\scriptsize$[-5.7, +2.0]$} \\
Gemini-3.1 & 82.1 & 85.1 & 82.8 & 82.6 & $+3.0$\,{\scriptsize$[-0.1, +6.2]$} & $+1.0$\,{\scriptsize$[-2.0, +4.1]$} & $-3.2$\,{\scriptsize$[-7.4, +0.7]$} \\
MiMo-2.6-Pro & 76.5 & 75.0 & 75.9 & 76.9 & $-0.9$\,{\scriptsize$[-4.5, +2.8]$} & $-0.1$\,{\scriptsize$[-4.0, +3.6]$} & $+0.7$\,{\scriptsize$[-4.0, +5.3]$} \\
MiMo-2.6-Flash & 71.0 & 78.1 & 73.7 & 70.5 & $\mathbf{+7.7}$\,{\scriptsize$\boldsymbol{[+3.0, +12.2]}$} & $+2.6$\,{\scriptsize$[-2.4, +7.6]$} & $\mathbf{-9.4}$\,{\scriptsize$\boldsymbol{[-16.5, -2.3]}$} \\
Doubao-2.0 & 75.1 & 72.4 & 71.1 & 70.0 & $-2.6$\,{\scriptsize$[-6.3, +0.9]$} & $\mathbf{-3.9}$\,{\scriptsize$\boldsymbol{[-7.8, 0.0]}$} & $+1.5$\,{\scriptsize$[-3.8, +6.8]$} \\
Muse-1.3 & 64.7 & 65.8 & 64.2 & 62.7 & $+1.0$\,{\scriptsize$[-3.6, +5.2]$} & $-0.7$\,{\scriptsize$[-5.3, +3.8]$} & $-2.6$\,{\scriptsize$[-9.1, +4.3]$} \\
\midrule
Qwen-2.5 & 31.3 & 30.1 & 29.9 & 29.7 & $-0.6$\,{\scriptsize$[-4.2, +3.3]$} & $-0.6$\,{\scriptsize$[-4.8, +3.5]$} & $-0.9$\,{\scriptsize$[-6.0, +4.5]$} \\
MiMo-Audio & 22.5 & 20.5 & 22.2 & 21.4 & $-2.0$\,{\scriptsize$[-5.7, +1.8]$} & $-0.2$\,{\scriptsize$[-3.8, +3.2]$} & $+1.2$\,{\scriptsize$[-3.5, +6.1]$} \\
\bottomrule
\end{tabular}
\caption{Task accuracy per cell and effects of frustrated voice and wording with 95\% CIs, macro-averaged over task families. Bold: CI excludes zero.}
\label{tab:app_effects}
\end{table*}

\begin{table*}[t]
\centering
\scriptsize
\setlength{\tabcolsep}{3pt}
\begin{adjustbox}{max width=\textwidth}\begin{tabular}{l l rrrr lll}
\toprule
Model & Family & BN & BV & AN & AV & Vocal $\Delta_v$ & Wording $\Delta_w$ & Interaction \\
\midrule
Gemini-3.8 & MMLU-Pro & 89.4 & 88.2 & 84.7 & 84.7 & $-1.2$\,{\scriptsize$[-5.9, +2.4]$} & $\mathbf{-4.7}$\,{\scriptsize$[-9.4, -1.2]$} & $+1.2$\,{\scriptsize$[-5.9, +8.2]$} \\
 & GPQA-Diamond & 87.5 & 88.9 & 87.7 & 90.1 & $+2.5$\,{\scriptsize$[-3.8, +10.0]$} & $+1.3$\,{\scriptsize$[-6.3, +10.1]$} & $-3.9$\,{\scriptsize$[-14.3, +5.2]$} \\
 & IFEval & 78.7 & 78.7 & 80.9 & 75.0 & $0.0$\,{\scriptsize$[-5.6, +5.6]$} & $+2.2$\,{\scriptsize$[-5.6, +10.1]$} & $-5.7$\,{\scriptsize$[-14.8, +2.3]$} \\
 & MATH-500 & 81.1 & 82.1 & 80.0 & 82.1 & $+1.1$\,{\scriptsize$[-2.1, +5.3]$} & $-1.1$\,{\scriptsize$[-5.3, +2.1]$} & $+1.1$\,{\scriptsize$[-3.2, +5.3]$} \\
Gemini-3.1 & MMLU-Pro & 88.2 & 89.4 & 86.9 & 85.9 & $+1.2$\,{\scriptsize$[-3.5, +5.9]$} & $-1.2$\,{\scriptsize$[-7.1, +3.6]$} & $-1.2$\,{\scriptsize$[-8.3, +6.0]$} \\
 & GPQA-Diamond & 79.3 & 86.7 & 88.0 & 89.3 & $+7.4$\,{\scriptsize$[-1.2, +16.0]$} & $+9.8$\,{\scriptsize$[0.0, +19.5]$} & $-7.4$\,{\scriptsize$[-18.5, +3.7]$} \\
 & IFEval & 80.9 & 82.0 & 77.5 & 76.4 & $+1.1$\,{\scriptsize$[-4.5, +6.7]$} & $-3.4$\,{\scriptsize$[-9.0, +2.2]$} & $-2.2$\,{\scriptsize$[-11.2, +6.7]$} \\
 & MATH-500 & 80.0 & 82.1 & 78.9 & 78.9 & $+2.1$\,{\scriptsize$[-2.1, +6.3]$} & $-1.1$\,{\scriptsize$[-3.2, 0.0]$} & $-2.1$\,{\scriptsize$[-6.3, +2.1]$} \\
MiMo-2.6-Pro & MMLU-Pro & 86.4 & 82.1 & 82.5 & 84.8 & $-2.5$\,{\scriptsize$[-7.5, +2.5]$} & $-5.2$\,{\scriptsize$[-11.7, 0.0]$} & $+4.1$\,{\scriptsize$[-4.1, +12.2]$} \\
 & GPQA-Diamond & 78.5 & 71.2 & 75.0 & 76.8 & $-6.8$\,{\scriptsize$[-16.4, +2.7]$} & $-1.3$\,{\scriptsize$[-11.8, +9.2]$} & $+8.7$\,{\scriptsize$[-2.9, +20.3]$} \\
 & IFEval & 68.5 & 73.6 & 73.8 & 73.6 & $+5.7$\,{\scriptsize$[-2.3, +13.8]$} & $+7.1$\,{\scriptsize$[0.0, +14.3]$} & $\mathbf{-10.0}$\,{\scriptsize$[-20.0, -1.2]$} \\
 & MATH-500 & 72.6 & 73.1 & 72.3 & 72.3 & $0.0$\,{\scriptsize$[-5.4, +5.4]$} & $-1.1$\,{\scriptsize$[-6.4, +4.3]$} & $0.0$\,{\scriptsize$[-7.6, +7.6]$} \\
MiMo-2.6-Flash & MMLU-Pro & 75.3 & 83.3 & 80.0 & 80.8 & $+6.8$\,{\scriptsize$[-1.4, +16.4]$} & $+1.4$\,{\scriptsize$[-8.5, +11.3]$} & $-4.7$\,{\scriptsize$[-18.8, +7.8]$} \\
 & GPQA-Diamond & 69.2 & 79.2 & 67.1 & 68.8 & $+10.4$\,{\scriptsize$[-1.5, +22.4]$} & $-1.4$\,{\scriptsize$[-14.1, +11.3]$} & $-5.3$\,{\scriptsize$[-24.6, +14.0]$} \\
 & IFEval & 71.3 & 75.6 & 70.6 & 67.9 & $+7.5$\,{\scriptsize$[-1.2, +16.2]$} & $0.0$\,{\scriptsize$[-8.4, +8.4]$} & $-10.8$\,{\scriptsize$[-24.3, +1.4]$} \\
 & MATH-500 & 68.2 & 74.4 & 76.9 & 64.5 & $+5.9$\,{\scriptsize$[0.0, +11.8]$} & $\mathbf{+10.6}$\,{\scriptsize$[+1.2, +20.0]$} & $\mathbf{-16.9}$\,{\scriptsize$[-26.5, -7.2]$} \\
Doubao-2.0 & MMLU-Pro & 86.7 & 81.9 & 81.0 & 79.0 & $-4.9$\,{\scriptsize$[-11.0, 0.0]$} & $-6.1$\,{\scriptsize$[-12.2, 0.0]$} & $+2.5$\,{\scriptsize$[-5.1, +11.4]$} \\
 & GPQA-Diamond & 69.9 & 61.0 & 65.5 & 63.9 & $-8.6$\,{\scriptsize$[-17.3, 0.0]$} & $-3.6$\,{\scriptsize$[-13.3, +4.8]$} & $+7.5$\,{\scriptsize$[-6.2, +21.2]$} \\
 & IFEval & 77.5 & 74.2 & 70.8 & 69.7 & $-3.4$\,{\scriptsize$[-10.1, +3.4]$} & $-6.7$\,{\scriptsize$[-14.6, 0.0]$} & $+2.2$\,{\scriptsize$[-6.7, +11.2]$} \\
 & MATH-500 & 66.3 & 72.6 & 67.4 & 67.4 & $+6.3$\,{\scriptsize$[-1.1, +13.7]$} & $+1.1$\,{\scriptsize$[-5.3, +7.4]$} & $-6.3$\,{\scriptsize$[-16.8, +4.2]$} \\
Muse-1.3 & MMLU-Pro & 63.1 & 63.1 & 64.3 & 65.5 & $0.0$\,{\scriptsize$[-10.7, +10.7]$} & $+1.2$\,{\scriptsize$[-8.3, +11.9]$} & $+1.2$\,{\scriptsize$[-13.1, +15.5]$} \\
 & GPQA-Diamond & 63.9 & 72.0 & 70.7 & 68.7 & $+7.3$\,{\scriptsize$[-1.2, +17.1]$} & $+6.1$\,{\scriptsize$[-3.7, +15.9]$} & $-9.9$\,{\scriptsize$[-23.5, +3.7]$} \\
 & IFEval & 70.8 & 65.2 & 60.7 & 61.8 & $-5.6$\,{\scriptsize$[-13.5, +2.2]$} & $\mathbf{-10.1}$\,{\scriptsize$[-18.0, -2.2]$} & $+6.7$\,{\scriptsize$[-5.6, +19.1]$} \\
 & MATH-500 & 61.1 & 63.2 & 61.1 & 54.7 & $+2.1$\,{\scriptsize$[-5.3, +9.5]$} & $0.0$\,{\scriptsize$[-8.4, +8.4]$} & $-8.4$\,{\scriptsize$[-18.9, +2.1]$} \\
\midrule
Qwen-2.5 & MMLU-Pro & 34.9 & 37.3 & 32.5 & 34.9 & $+2.4$\,{\scriptsize$[-4.8, +9.6]$} & $-2.4$\,{\scriptsize$[-8.5, +3.6]$} & $0.0$\,{\scriptsize$[-8.5, +9.6]$} \\
 & GPQA-Diamond & 30.1 & 25.3 & 26.5 & 26.5 & $-4.8$\,{\scriptsize$[-13.3, +2.4]$} & $-3.6$\,{\scriptsize$[-13.3, +6.0]$} & $+4.8$\,{\scriptsize$[-6.0, +15.7]$} \\
 & IFEval & 38.4 & 39.1 & 39.3 & 42.0 & $0.0$\,{\scriptsize$[-7.1, +7.1]$} & $+1.2$\,{\scriptsize$[-7.0, +10.5]$} & $+1.2$\,{\scriptsize$[-9.6, +12.0]$} \\
 & MATH-500 & 21.8 & 18.7 & 21.3 & 15.4 & $0.0$\,{\scriptsize$[-7.1, +7.1]$} & $+2.4$\,{\scriptsize$[-4.7, +9.4]$} & $-9.6$\,{\scriptsize$[-20.5, +1.2]$} \\
MiMo-Audio & MMLU-Pro & 29.4 & 25.9 & 27.1 & 24.7 & $-3.5$\,{\scriptsize$[-11.8, +4.7]$} & $-2.4$\,{\scriptsize$[-10.6, +5.9]$} & $+1.2$\,{\scriptsize$[-10.6, +12.9]$} \\
 & GPQA-Diamond & 21.4 & 20.2 & 23.8 & 28.6 & $-1.2$\,{\scriptsize$[-10.7, +8.3]$} & $+2.4$\,{\scriptsize$[-4.8, +9.5]$} & $+6.0$\,{\scriptsize$[-6.0, +19.0]$} \\
 & IFEval & 33.7 & 34.8 & 34.8 & 30.3 & $+1.1$\,{\scriptsize$[-4.5, +6.7]$} & $+1.1$\,{\scriptsize$[-5.6, +7.9]$} & $-5.6$\,{\scriptsize$[-14.6, +2.2]$} \\
 & MATH-500 & 5.3 & 1.1 & 3.2 & 2.1 & $-4.2$\,{\scriptsize$[-9.5, 0.0]$} & $-2.1$\,{\scriptsize$[-7.4, +3.2]$} & $+3.2$\,{\scriptsize$[-1.1, +8.4]$} \\
\bottomrule
\end{tabular}\end{adjustbox}
\caption{Accuracy (\%) by task family and within-family effects (pp, 95\% CI). Bold: CI excludes zero.}
\label{tab:app_accbench}
\end{table*}

\paragraph{Paired bootstrap.} All accuracy and perception contrasts are within item. For a contrast between two cells, we compute each item's paired difference, average within each task family, and macro-average the 4 family means. CIs come from 5{,}000 bootstrap resamples that resample items with replacement within each family, which keeps the family composition fixed, and we report percentile intervals. Single-family results use the same procedure within one family.

\paragraph{Perception statistics.} $S_v$, $S_w$, and the cell contrasts of the probe use the same paired bootstrap. Voice and text accuracy are the shares of parsed labels that match the vocal or the wording condition. RAVDESS hit and false-alarm rates use Clopper--Pearson intervals, and $d'$ uses a log-linear correction that adds 0.5 to hit and false-alarm counts and 1 to the number of trials. The paired-clip probe uses an exact binomial test against 50\%, and the excerpt probe an item bootstrap stratified by family with exact McNemar tests.

\paragraph{Expression statistics.} SER differences are paired by item and wording. CIs come from 5{,}000 item resamples, and $d_z$ is the mean paired difference divided by its standard deviation. Arousal quartiles are formed from each item's arousal contrast averaged over both wordings.

\subsection{Regimes}
\label{app:stats-regimes}

\begin{figure}[H]
\centering
\includegraphics[width=\columnwidth]{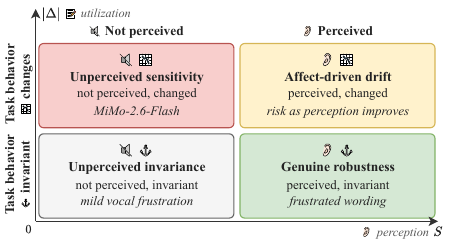}
\caption{The 4 regimes of robustness to user frustration. Output invariance is informative only when the frustration is perceived. Italics mark where the evaluated models fall.}
\label{fig:regimes}
\end{figure}

\begin{figure}[H]
\centering
\includegraphics[width=\columnwidth]{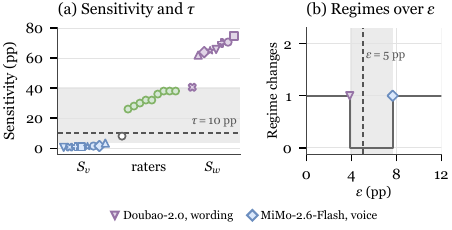}
\caption{Robustness of the regime thresholds. (a) Sensitivity of the 8 models, marked as in Figure~\ref{fig:plane}, and of the 10 raters, green when significant. (b) Regime changes over $\varepsilon$. Shading marks the threshold values that keep all regimes.}
\label{fig:thresholds}
\end{figure}

Figure~\ref{fig:regimes} shows the 4 regimes, and Figure~\ref{fig:plane} draws their thresholds. We calibrate $\tau$ on the human listeners of Section~\ref{sec:validation}, who make the same binary tone judgment. Their sensitivity, the hit rate minus the false-alarm rate, is 30.6 pp pooled, 26 to 38 pp for the 9 raters whose accuracy is individually significant, and 8 pp for the remaining rater, so $\tau=10$ pp counts exactly the 9 significant raters as perceiving the voice. The largest acoustic sensitivity of any model, 3.4 pp with an upper CI bound of 5.6 pp, falls below even the remaining rater. Figure~\panelref{fig:thresholds}{a} shows that the regimes of all 8 models stay the same for every $\tau$ between 3.4 and 40.6 pp. The margin $\varepsilon$ separates practically relevant effects from small ones that the CI test alone would flag. Figure~\panelref{fig:thresholds}{b} shows that at any $\varepsilon$, at most 1 model changes regime. Below 3.9 pp, the wording effect of Doubao-2.0 counts as affect-driven drift, and above 7.7 pp, the vocal effect of MiMo-2.6-Flash counts as unperceived invariance. In both cases, the vocal channel of every model remains unperceived. Table~\ref{tab:app_sig_holm} shows that Holm correction over the 24 accuracy contrasts leaves only MiMo-2.6-Flash's vocal effect significant, matching the $\varepsilon$ rule.

\subsection{Significance}
\label{app:stats-sig}

Table~\ref{tab:app_sig_overview} maps each claim type to its test, resampling unit, and multiplicity handling. For accuracy, we add two-sided bootstrap $p$-values to the paired bootstrap of the accuracy CIs and apply Holm correction over the 24 model--contrast tests. Table~\ref{tab:app_sig_holm} reports the estimates with raw and corrected $p$-values.

\begin{table}[H]
\centering
\small
\setlength{\tabcolsep}{3.5pt}
\begin{adjustbox}{max width=\columnwidth}\begin{tabular}{@{}llll@{}}
\toprule
Contrast & Test & Unit & Multiplicity \\
\midrule
Accuracy $\Delta_v$, $\Delta_w$, $I$ & Bootstrap CI; two-sided $p$ & Items/family & Holm over 24 \\
Perception $S_v$, $S_w$ & Paired bootstrap CI & Items/family & --- \\
RAVDESS control & Clopper--Pearson; $d'$ & Clips & --- \\
Human listening & Item bootstrap; binomial & Items / trials & Per-rater $\alpha=0.05$ \\
Hidden-state $A$ & Item bootstrap CI & Items & Pre-specified layers \\
Expression $E$ & Paired bootstrap CI; $d_z$ & Items & --- \\
Paired-clip probe & Exact binomial vs.\ 50\% & Trials & --- \\
Excerpt probe & Exact McNemar & Items & --- \\
\bottomrule
\end{tabular}\end{adjustbox}
\caption{Tests behind the main claims. Accuracy $p$-values are Holm-corrected in Table~\ref{tab:app_sig_holm}.}
\label{tab:app_sig_overview}
\end{table}

\begin{table}[H]
\centering
\small
\setlength{\tabcolsep}{3.5pt}
\begin{adjustbox}{max width=\columnwidth}\begin{tabular}{l l r r r}
\toprule
Model & Effect & Est.\ (pp) & $p$ & $p_{\mathrm{Holm}}$ \\
\midrule
Gemini-3.8 & $\Delta_v$ & +0.6 & 0.6524 & 1.0000 \\
Gemini-3.8 & $\Delta_w$ & -0.6 & 0.7332 & 1.0000 \\
Gemini-3.8 & $I$ & -1.8 & 0.3548 & 1.0000 \\
\midrule
Gemini-3.1 & $\Delta_v$ & +3.0 & 0.0592 & 1.0000 \\
Gemini-3.1 & $\Delta_w$ & +1.0 & 0.5044 & 1.0000 \\
Gemini-3.1 & $I$ & -3.2 & 0.1080 & 1.0000 \\
\midrule
MiMo-2.6-Pro & $\Delta_v$ & -0.9 & 0.6268 & 1.0000 \\
MiMo-2.6-Pro & $\Delta_w$ & -0.1 & 0.9652 & 1.0000 \\
MiMo-2.6-Pro & $I$ & +0.7 & 0.7568 & 1.0000 \\
\midrule
MiMo-2.6-Flash & $\Delta_v$ & \textbf{+7.7} & \textbf{0.0008} & \textbf{0.0192} \\
MiMo-2.6-Flash & $\Delta_w$ & +2.6 & 0.3148 & 1.0000 \\
MiMo-2.6-Flash & $I$ & -9.4 & 0.0092 & 0.2116 \\
\midrule
Doubao-2.0 & $\Delta_v$ & -2.6 & 0.1400 & 1.0000 \\
Doubao-2.0 & $\Delta_w$ & -3.9 & 0.0488 & 1.0000 \\
Doubao-2.0 & $I$ & +1.5 & 0.5820 & 1.0000 \\
\midrule
Muse-1.3 & $\Delta_v$ & +1.0 & 0.6880 & 1.0000 \\
Muse-1.3 & $\Delta_w$ & -0.7 & 0.7372 & 1.0000 \\
Muse-1.3 & $I$ & -2.6 & 0.4536 & 1.0000 \\
\midrule
Qwen-2.5 & $\Delta_v$ & -0.6 & 0.7324 & 1.0000 \\
Qwen-2.5 & $\Delta_w$ & -0.6 & 0.7536 & 1.0000 \\
Qwen-2.5 & $I$ & -0.9 & 0.7356 & 1.0000 \\
\midrule
MiMo-Audio & $\Delta_v$ & -2.0 & 0.3120 & 1.0000 \\
MiMo-Audio & $\Delta_w$ & -0.2 & 0.9088 & 1.0000 \\
MiMo-Audio & $I$ & +1.2 & 0.6488 & 1.0000 \\
\bottomrule
\end{tabular}\end{adjustbox}
\caption{Two-sided bootstrap $p$-values for the 24 accuracy effects, with Holm correction. Bold: $p_{\mathrm{Holm}}<0.05$.}
\label{tab:app_sig_holm}
\end{table}

\begin{table*}[t]
\centering
\small
\setlength{\tabcolsep}{3pt}
\begin{tabular}{l rrrr lll rr}
\toprule
& \multicolumn{4}{c}{$P(\textit{frustrated})$ (\%)} & \multicolumn{3}{c}{Contrast (pp) [95\% CI]} & \multicolumn{2}{c}{Match (\%)} \\
\cmidrule(lr){2-5}\cmidrule(lr){6-8}\cmidrule(lr){9-10}
Model & BN & BV & AN & AV & BV$-$BN & $S_v$ & $S_w$ & Voice & Text \\
\midrule
Gemini-3.8 & 0.5 & 0.8 & 70.0 & 72.5 & $+0.3$\,{\scriptsize$[0.0, +0.8]$} & $+1.4$\,{\scriptsize$[-0.2, +3.0]$} & $\mathbf{+71}$\,{\scriptsize$[+66, +75]$} & 50.7 & 85.3 \\
Gemini-3.1 & 1.1 & 1.1 & 74.8 & 76.7 & $0.0$\,{\scriptsize$[-1.7, +1.6]$} & $+0.9$\,{\scriptsize$[-0.8, +2.6]$} & $\mathbf{+75}$\,{\scriptsize$[+70, +79]$} & 50.5 & 87.3 \\
MiMo-2.6-Pro & 0.8 & 1.6 & 60.2 & 66.1 & $+0.8$\,{\scriptsize$[0.0, +1.9]$} & $\mathbf{+3.4}$\,{\scriptsize$[+1.2, +5.5]$} & $\mathbf{+62}$\,{\scriptsize$[+57, +66]$} & 51.7 & 81.0 \\
MiMo-2.6-Flash & 0.6 & 1.7 & 64.3 & 65.1 & $\mathbf{+1.1}$\,{\scriptsize$[+0.3, +2.3]$} & $+1.4$\,{\scriptsize$[-0.9, +3.8]$} & $\mathbf{+64}$\,{\scriptsize$[+59, +68]$} & 50.4 & 81.8 \\
Doubao-2.0 & 0.3 & 0.5 & 65.7 & 66.4 & $+0.3$\,{\scriptsize$[0.0, +0.8]$} & $+0.5$\,{\scriptsize$[-1.0, +2.1]$} & $\mathbf{+66}$\,{\scriptsize$[+61, +70]$} & 50.3 & 82.9 \\
Muse-1.3 & 0.0 & 0.0 & 69.3 & 71.0 & $0.0$\,{\scriptsize$[0.0, 0.0]$} & $+0.9$\,{\scriptsize$[-1.1, +2.9]$} & $\mathbf{+70}$\,{\scriptsize$[+66, +74]$} & 50.4 & 84.9 \\
\midrule
Qwen-2.5 & 0.3 & 0.0 & 39.9 & 41.4 & $-0.3$\,{\scriptsize$[-0.8, 0.0]$} & $+0.7$\,{\scriptsize$[-0.7, +2.2]$} & $\mathbf{+41}$\,{\scriptsize$[+36, +45]$} & 50.3 & 70.2 \\
MiMo-Audio & 2.5 & 3.1 & 66.9 & 68.6 & $+0.5$\,{\scriptsize$[-0.6, +1.7]$} & $+1.1$\,{\scriptsize$[-0.2, +2.4]$} & $\mathbf{+65}$\,{\scriptsize$[+60, +70]$} & 50.6 & 82.4 \\
\bottomrule
\end{tabular}
\caption{Tone probe on the task clips: rate of \emph{frustrated} labels per cell, perception contrasts with 95\% CIs, and label agreement with the vocal and wording conditions. Bold: CI excludes zero.}
\label{tab:perception_full}
\end{table*}

\begin{table*}[t]
\centering
\small
\setlength{\tabcolsep}{4pt}
\begin{adjustbox}{max width=\textwidth}\begin{tabular}{l rr rr rr rr r}
\toprule
& \multicolumn{2}{c}{Neutral voice} & \multicolumn{2}{c}{Frustrated voice} & \multicolumn{2}{c}{Neutral wording} & \multicolumn{2}{c}{Frustrated wording} & \\
\cmidrule(lr){2-3}\cmidrule(lr){4-5}\cmidrule(lr){6-7}\cmidrule(lr){8-9}
Model & neu. & frus. & neu. & frus. & neu. & frus. & neu. & frus. & Unparsed \\
\midrule
Gemini-3.8 & 457 & 249 & 447 & 259 & 701 & 5 & 203 & 503 & 0 \\
Gemini-3.1 & 438 & 268 & 431 & 275 & 698 & 8 & 171 & 535 & 0 \\
MiMo-2.6-Pro & 490 & 216 & 466 & 240 & 697 & 9 & 259 & 447 & 0 \\
MiMo-2.6-Flash & 470 & 225 & 469 & 236 & 692 & 8 & 247 & 453 & 12 \\
Doubao-2.0 & 473 & 233 & 469 & 237 & 703 & 3 & 239 & 467 & 0 \\
Muse-1.3 & 460 & 244 & 455 & 249 & 703 & 0 & 212 & 493 & 4 \\
\midrule
Qwen-2.5 & 564 & 141 & 560 & 146 & 705 & 1 & 419 & 286 & 1 \\
MiMo-Audio & 461 & 245 & 453 & 253 & 686 & 20 & 228 & 478 & 0 \\
\bottomrule
\end{tabular}\end{adjustbox}
\caption{Confusion counts of the binary tone probe by vocal and by wording condition.}
\label{tab:app_confusion}
\end{table*}

\begin{table*}[t]
\centering
\small
\setlength{\tabcolsep}{3.5pt}
\begin{adjustbox}{max width=\textwidth}\begin{tabular}{l rrrr l rrrr l}
\toprule
& \multicolumn{5}{c}{Neutral wording: BV$-$BN (pp)} & \multicolumn{5}{c}{Frustrated wording: AV$-$AN (pp)} \\
\cmidrule(lr){2-6}\cmidrule(lr){7-11}
Model & Q1 & Q2 & Q3 & Q4 & Q4$-$Q1 [95\% CI] & Q1 & Q2 & Q3 & Q4 & Q4$-$Q1 [95\% CI] \\
\midrule
Gemini-3.8 & 0.0 & 0.0 & +1.1 & 0.0 & $0.0$\,{\scriptsize$[0.0, 0.0]$} & -4.5 & +3.4 & +2.3 & +9.1 & $\mathbf{+13.6}$\,{\scriptsize$[+4.6, +22.7]$} \\
Gemini-3.1 & -3.4 & +1.1 & 0.0 & +2.3 & $\mathbf{+5.6}$\,{\scriptsize$[+1.1, +11.3]$} & -1.1 & +5.7 & +1.1 & +2.3 & $+3.4$\,{\scriptsize$[-4.5, +11.3]$} \\
MiMo-2.6-Pro & +1.1 & 0.0 & +1.1 & +1.1 & $0.0$\,{\scriptsize$[-3.4, +3.4]$} & -3.4 & +9.1 & +8.0 & +10.2 & $\mathbf{+13.6}$\,{\scriptsize$[+2.3, +24.9]$} \\
MiMo-2.6-Flash & +1.1 & +1.2 & +2.3 & 0.0 & $-1.1$\,{\scriptsize$[-3.4, 0.0]$} & +3.5 & +1.1 & 0.0 & +2.3 & $-1.2$\,{\scriptsize$[-12.7, +10.4]$} \\
Doubao-2.0 & +1.1 & 0.0 & 0.0 & 0.0 & $-1.1$\,{\scriptsize$[-3.4, 0.0]$} & -2.2 & +1.1 & +1.1 & +3.4 & $+5.7$\,{\scriptsize$[-2.3, +13.6]$} \\
Muse-1.3 & 0.0 & 0.0 & 0.0 & 0.0 & $0.0$\,{\scriptsize$[0.0, 0.0]$} & -2.2 & +4.5 & -2.3 & +6.8 & $+9.1$\,{\scriptsize$[-1.1, +19.3]$} \\
\midrule
Qwen-2.5 & 0.0 & 0.0 & 0.0 & -1.1 & $-1.1$\,{\scriptsize$[-3.4, 0.0]$} & -5.7 & +3.4 & +3.4 & +5.7 & $\mathbf{+11.4}$\,{\scriptsize$[+4.5, +19.3]$} \\
MiMo-Audio & -1.1 & 0.0 & +3.4 & 0.0 & $+1.1$\,{\scriptsize$[0.0, +3.4]$} & +3.4 & -2.3 & +3.4 & +2.3 & $-1.1$\,{\scriptsize$[-7.9, +5.7]$} \\
\bottomrule
\end{tabular}\end{adjustbox}
\caption{Vocal effect on the tone probe by arousal-contrast quartile. Bold: CI excludes zero.}
\label{tab:app_quartiles}
\end{table*}

\begin{table*}[t]
\centering
\small
\setlength{\tabcolsep}{2.1pt}
\begin{tabular}{l rrlrrr rrlr rr}
\toprule
& \multicolumn{6}{c}{Core block (neutral wording)} & \multicolumn{4}{c}{Optional block (frustrated wording)} & \multicolumn{2}{c}{Full clips} \\
\cmidrule(lr){2-7}\cmidrule(lr){8-11}\cmidrule(lr){12-13}
Model & BN & BV & BV$-$BN [95\% CI] & Acc. & $d'$ & Catch & AN & AV & AV$-$AN [95\% CI] & AN$-$BN & BN & BV \\
\midrule
Gemini-3.8 & 4.0 & 0.0 & $-4.0$\,{\scriptsize$[-10.0, 0.0]$} & 48.0 & $-0.68$ & 5/6 & 62.0 & 65.3 & $+2.0$\,{\scriptsize$[0.0, +6.1]$} & $+58$ & 2.0 & 2.0 \\
Gemini-3.1 & 0.0 & 0.0 & $0.0$\,{\scriptsize$[0.0, 0.0]$} & 50.0 & $0.00$ & 4/6 & 58.0 & 64.0 & $+6.0$\,{\scriptsize$[-2.0, +14.0]$} & $+58$ & 0.0 & 2.0 \\
MiMo-2.6-Pro & 4.0 & 0.0 & $-4.0$\,{\scriptsize$[-10.0, 0.0]$} & 48.0 & $-0.68$ & 5/6 & 62.0 & 66.0 & $+4.0$\,{\scriptsize$[0.0, +10.0]$} & $+58$ & 2.0 & 2.0 \\
MiMo-2.6-Flash & 2.0 & 2.0 & $0.0$\,{\scriptsize$[-6.0, +6.0]$} & 50.0 & $0.00$ & 6/6 & 60.0 & 62.0 & $+2.0$\,{\scriptsize$[-6.0, +10.0]$} & $+58$ & 0.0 & 4.0 \\
Doubao-2.0 & 2.0 & 2.0 & $0.0$\,{\scriptsize$[0.0, 0.0]$} & 50.0 & $0.00$ & 6/6 & 56.0 & 52.0 & $-4.0$\,{\scriptsize$[-12.0, +4.0]$} & $+54$ & 2.0 & 2.0 \\
Muse-1.3 & 0.0 & 0.0 & $0.0$\,{\scriptsize$[0.0, 0.0]$} & 50.0 & $0.00$ & 3/6 & 58.0 & 64.0 & $+6.0$\,{\scriptsize$[0.0, +14.0]$} & $+58$ & 0.0 & 0.0 \\
\midrule
Qwen-2.5 & 0.0 & 0.0 & $0.0$\,{\scriptsize$[0.0, 0.0]$} & 50.0 & $0.00$ & 5/6 & 42.0 & 48.0 & $+6.0$\,{\scriptsize$[0.0, +12.0]$} & $+42$ & 0.0 & 0.0 \\
MiMo-Audio & 2.0 & 0.0 & $-2.0$\,{\scriptsize$[-6.0, 0.0]$} & 49.0 & $-0.44$ & 6/6 & 62.0 & 60.0 & $-2.0$\,{\scriptsize$[-6.0, 0.0]$} & $+60$ & 2.0 & 0.0 \\
\midrule
Humans & 25.2 & 55.8 & $\mathbf{+30.6}$\,{\scriptsize$[+22.2, +39.0]$} & 65.3 & $0.81$ & 60/60 & 38.2 & 63.0 & $\mathbf{+24.8}$\,{\scriptsize$[+16.6, +33.2]$} & $+13$ & -- & -- \\
\bottomrule
\end{tabular}
\caption{Tone probe on the 15-s listening excerpts, with RAVDESS catch trials and the full clips of the same 50 items. Humans: 10 raters, pooled over judgments. --: not measured. Bold: CI excludes zero.}
\label{tab:excerpt_probe}
\end{table*}

\begin{table*}[t]
\centering
\small
\setlength{\tabcolsep}{4pt}
\begin{tabular*}{\textwidth}{@{\extracolsep{\fill}} l rrrr rrrr @{}}
\toprule
& \multicolumn{4}{c}{Acoustic sensitivity $S_v$ (pp)} & \multicolumn{4}{c}{Lexical sensitivity $S_w$ (pp)} \\
\cmidrule(lr){2-5}\cmidrule(lr){6-9}
Model & MMLU-Pro & GPQA & IFEval & MATH-500 & MMLU-Pro & GPQA & IFEval & MATH-500 \\
\midrule
Gemini-3.8 & $+0.6$ & $+0.6$ & $+3.4$ & $+1.1$ & $+69$ & $+70$ & $+79$ & $+65$ \\
Gemini-3.1 & $-2.4$ & $+1.2$ & $+2.2$ & $+2.6$ & $+74$ & $+69$ & $+84$ & $+71$ \\
MiMo-2.6-Pro & $+1.8$ & $+3.0$ & $+3.9$ & $+4.7$ & $+54$ & $+60$ & $+74$ & $+61$ \\
MiMo-2.6-Flash & $-3.9$ & $+1.0$ & $+4.5$ & $+2.2$ & $+60$ & $+62$ & $+72$ & $+60$ \\
Doubao-2.0 & $0.0$ & $-2.4$ & $+2.8$ & $+1.6$ & $+64$ & $+63$ & $+71$ & $+65$ \\
Muse-1.3 & $+2.9$ & $-0.2$ & $+1.7$ & $-1.1$ & $+71$ & $+75$ & $+71$ & $+63$ \\
\midrule
Qwen-2.5 & $0.0$ & $0.0$ & $+0.9$ & $+1.6$ & $+33$ & $+46$ & $+46$ & $+36$ \\
MiMo-Audio & $0.0$ & $+0.6$ & $+1.1$ & $+2.6$ & $+62$ & $+67$ & $+70$ & $+61$ \\
\bottomrule
\end{tabular*}
\caption{Acoustic and lexical sensitivity of the tone probe by task family.}
\label{tab:app_perception_family}
\end{table*}

\section{Accuracy Details}
\label{app:tables}

\begin{figure}[H]
\centering
\includegraphics[width=0.94\columnwidth]{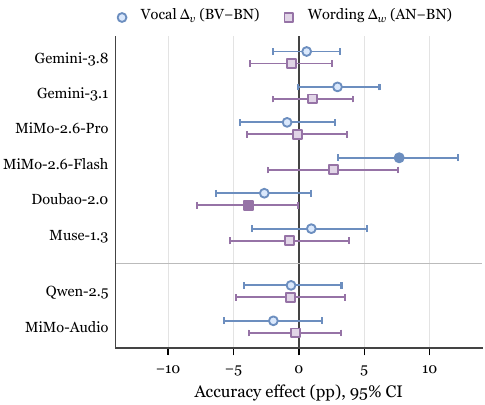}
\caption{Accuracy effects of frustrated voice and wording with 95\% CIs. Filled markers: CI excludes zero.}
\label{fig:effects}
\end{figure}

Table~\ref{tab:app_effects} gives the cell accuracies and the effects behind Table~\ref{tab:main_effects} with 95\% CIs, and Table~\ref{tab:app_accbench} breaks them down by task family. Figure~\ref{fig:effects} plots the macro-averaged effects. MiMo-2.6-Flash's vocal effect has the same sign in all 4 families.

\paragraph{Details for \obsref{4}.} MiMo-2.6-Flash's vocal effect of $+7.7$ pp [$+3.0$, $+12.2$] disappears under frustrated wording, which yields an interaction of $-9.4$ pp [$-16.5$, $-2.3$]. Its tone probe does not register the voice ($S_v=1.4$ pp [$-0.9$, $+3.8$], excerpt BV$-$BN $=0$).

\section{Perception Details}
\label{app:perception}

Table~\ref{tab:perception_full} gives the label rates per cell and the perception contrasts with CIs, and Table~\ref{tab:app_confusion} the confusion counts behind them. For every model, the frustrated voice barely changes the labels, whereas frustrated wording moves 40--77\% of clips to \emph{frustrated}. Table~\ref{tab:app_quartiles} shows the probe of all 8 models by arousal quartile, where the vocal effect grows with arousal only under frustrated wording. Table~\ref{tab:app_perception_family} breaks both sensitivities down by task family, and the lexical sensitivity exceeds the acoustic one in every family for every model. Table~\ref{tab:app_ravdess} details the RAVDESS control, Tables~\ref{tab:extras} and~\ref{tab:app_variants} the open-label and paired-clip variants, and Table~\ref{tab:excerpt_probe} the probe on the 15-s listening excerpts.

\paragraph{Measures.} Voice and text match are the shares of parsed labels that agree with the vocal or the wording condition, with chance at 50\%. A keyword rule parses the returned label, and unparsed labels are missing. In the open-label probe, \emph{angry}, \emph{frustrated}, and \emph{sad} count as non-neutral. The RAVDESS control uses 96 strong-intensity angry and 96 neutral clips. Hits are angry clips labeled frustrated, false alarms (FA) are neutral clips labeled frustrated, and $d'$ uses a log-linear correction. Arousal quartiles have about 88 items each, with mean contrasts of $-0.007$, $+0.020$, $+0.042$, and $+0.089$, and Q4$-$Q1 differences use an item bootstrap. In the paired-clip probe, the frustrated clip is played first or second in 50 trials each.

\paragraph{Details for \obsref{2}.} Strong acted anger is labeled frustrated by Doubao-2.0 in 100\% of clips, MiMo-Audio in 96.9\%, MiMo-2.6-Pro in 95.8\%, MiMo-2.6-Flash in 94.8\%, Gemini-3.8 in 91.7\%, Gemini-3.1 in 75.0\%, and Qwen-2.5 in 72.6\% ($d'=3.2$--$5.1$). Muse-1.3 labels 20.8\% ($d'=1.8$). In the top arousal quartile (mean contrast $+0.089$, about 2.5 times the average), frustrated-voice clips with neutral wording are labeled frustrated in 0--5.7\% of cases. The open label set yields a non-neutral label for at most 2\% of BN or BV clips, and paired-clip accuracy is 50\% [40, 60] for Gemini-3.8 and 53\% [43, 63] for MiMo-2.6-Pro, with clip~2 chosen in 94\% and 77\% of trials. On the 15-s excerpts, core accuracy is 48--50\% and no BV$-$BN CI excludes zero.

\paragraph{Details for \obsref{3}.} Where $S_v$ departs from zero, the effect comes from the frustrated-wording cells. MiMo-2.6-Pro's $S_v$ of $+3.4$ pp [$+1.2$, $+5.5$] reflects AV$-$AN $=+5.9$ pp [$+1.7$, $+10.2$], while BV$-$BN is $+0.8$ pp, and its perception interaction of Eq.~\ref{eq:interaction} is $+5.1$ pp [$+0.8$, $+9.5$]. Table~\ref{tab:app_quartiles} shows that from the lowest to the highest arousal quartile, AV$-$AN rises by 13.6 pp for both Gemini-3.8 [$+4.6$, $+22.7$] and MiMo-2.6-Pro [$+2.3$, $+24.9$], and by 11.4 pp [$+4.5$, $+19.3$] for Qwen-2.5. Averaged over all 8 models, BV$-$BN stays near zero in every quartile, from $-0.1$ to $+1.0$ pp, with one significant rise of $+5.6$ pp [$+1.1$, $+11.3$] for Gemini-3.1. The amplifier holds at matched acoustic contrast. Among the clip pairs in the top quartile of arousal contrast, which averages 0.10 under both wordings, the voice raises the labels by 5.4 pp under frustrated wording and by 0.2 pp under neutral wording.

\begin{table}[H]
\centering
\small
\setlength{\tabcolsep}{3pt}
\begin{adjustbox}{max width=\columnwidth}\begin{tabular}{l r l r r}
\toprule
Model & Hits & Hit rate [95\% CI] & FA & $d'$ \\
\midrule
Gemini-3.8 & 88/96 & 91.7 {\scriptsize[84.2, 96.3]} & 0/96 & 3.92 \\
Gemini-3.1 & 72/96 & 75.0 {\scriptsize[65.1, 83.3]} & 0/96 & 3.23 \\
MiMo-2.6-Pro & 92/96 & 95.8 {\scriptsize[89.7, 98.9]} & 0/96 & 4.25 \\
MiMo-2.6-Flash & 91/96 & 94.8 {\scriptsize[88.3, 98.3]} & 0/96 & 4.15 \\
Doubao-2.0 & 96/96 & 100.0 {\scriptsize[96.2, 100.0]} & 0/96 & 5.13 \\
Muse-1.3 & 20/96 & 20.8 {\scriptsize[13.2, 30.3]} & 0/96 & 1.76 \\
\midrule
Qwen-2.5 & 69/95 & 72.6 {\scriptsize[62.5, 81.3]} & 0/96 & 3.16 \\
MiMo-Audio & 93/96 & 96.9 {\scriptsize[91.1, 99.4]} & 0/96 & 4.36 \\
\bottomrule
\end{tabular}\end{adjustbox}
\caption{RAVDESS positive control per model: hits, false alarms (FA), and $d'$.}
\label{tab:app_ravdess}
\end{table}

\begin{table}[H]
\centering
\small
\setlength{\tabcolsep}{4pt}
\begin{tabular}{l rr l}
\toprule
\multicolumn{4}{l}{\textit{(a) Open label set: $P(\textit{non-neutral})$ (\%)}} \\
Model & BN & BV & BV$-$BN [95\% CI] \\
\midrule
Gemini-3.8 & 1 & 0 & $-1$\,{\scriptsize$[-3, 0]$} \\
MiMo-2.6-Pro & 0 & 2 & $+2$\,{\scriptsize$[0, +5]$} \\
\bottomrule
\end{tabular}

\vspace{6pt}
\begin{tabular}{l r l r r}
\toprule
\multicolumn{5}{l}{\textit{(b) Paired clips: which sounds more frustrated?}} \\
Model & Acc. & 95\% CI & $p$ & Picks 2 \\
\midrule
Gemini-3.8 & 50 & [40, 60] & 1.00 & 94\% \\
MiMo-2.6-Pro & 53 & [43, 63] & .62 & 77\% \\
\bottomrule
\end{tabular}
\caption{Probe variants on the 100-item subset: (a) open label set and (b) two-clip forced choice. Label counts and position effects are in Table~\ref{tab:app_variants}.}
\label{tab:extras}
\end{table}

\begin{table}[H]
\centering
\small
\setlength{\tabcolsep}{3pt}
\begin{adjustbox}{max width=\columnwidth}\begin{tabular}{l l rrrrr}
\toprule
\multicolumn{7}{l}{\textit{(a) Open label set: label counts}} \\
Model & Clip & neu. & frus. & angry & sad & happy \\
\midrule
Gemini-3.8 & BN & 97 & 1 & 0 & 0 & 2 \\
Gemini-3.8 & BV & 96 & 0 & 0 & 0 & 4 \\
MiMo-2.6-Pro & BN & 96 & 0 & 0 & 0 & 4 \\
MiMo-2.6-Pro & BV & 95 & 1 & 1 & 0 & 3 \\
\bottomrule
\end{tabular}\end{adjustbox}

\vspace{6pt}
\begin{adjustbox}{max width=\columnwidth}\begin{tabular}{l rrr}
\toprule
\multicolumn{4}{l}{\textit{(b) Paired clips: accuracy by position of the frustrated clip}} \\
Model & First & Second & Picks 2 \\
\midrule
Gemini-3.8 & 6\% & 94\% & 94\% \\
MiMo-2.6-Pro & 26\% & 80\% & 77\% \\
\bottomrule
\end{tabular}\end{adjustbox}
\caption{Probe variants: (a) open-label counts and (b) paired-clip accuracy by position of the frustrated clip.}
\label{tab:app_variants}
\end{table}

\begin{table*}[t]
\centering
\small
\setlength{\tabcolsep}{4pt}
\begin{tabular}{l ll ll}
\toprule
& \multicolumn{2}{c}{Qwen-2.5} & \multicolumn{2}{c}{MiMo-Audio} \\
\cmidrule(lr){2-3}\cmidrule(lr){4-5}
Representation & Middle layer & Last layer & Middle layer & Last layer \\
\midrule
Audio side & $0.671$\,{\scriptsize$[0.619, 0.725]$} & $0.634$\,{\scriptsize$[0.584, 0.687]$} & $0.693$\,{\scriptsize$[0.643, 0.743]$} & $0.554$\,{\scriptsize$[0.507, 0.604]$} \\
Audio side, residualized & $0.630$\,{\scriptsize$[0.573, 0.684]$} & $0.575$\,{\scriptsize$[0.519, 0.628]$} & $0.651$\,{\scriptsize$[0.598, 0.705]$} & $0.560$\,{\scriptsize$[0.508, 0.618]$} \\
LLM, audio tokens & $0.601$\,{\scriptsize$[0.552, 0.647]$} & $0.567$\,{\scriptsize$[0.516, 0.620]$} & $0.542$\,{\scriptsize$[0.496, 0.590]$} & $0.520$\,{\scriptsize$[0.469, 0.573]$} \\
LLM, audio tokens, residualized & $0.657$\,{\scriptsize$[0.602, 0.710]$} & $0.609$\,{\scriptsize$[0.556, 0.662]$} & $0.589$\,{\scriptsize$[0.540, 0.636]$} & $0.601$\,{\scriptsize$[0.546, 0.655]$} \\
LLM, answer position & $0.619$\,{\scriptsize$[0.565, 0.676]$} & $0.611$\,{\scriptsize$[0.556, 0.668]$} & $0.538$\,{\scriptsize$[0.482, 0.596]$} & $0.539$\,{\scriptsize$[0.483, 0.595]$} \\
LLM, answer position, residualized & $0.586$\,{\scriptsize$[0.531, 0.640]$} & $0.557$\,{\scriptsize$[0.500, 0.613]$} & $0.513$\,{\scriptsize$[0.459, 0.572]$} & $0.548$\,{\scriptsize$[0.491, 0.605]$} \\
\midrule
Acoustic features (18) & \multicolumn{2}{c}{$0.667$\,{\scriptsize$[0.612, 0.722]$}} & \multicolumn{2}{c}{$0.669$\,{\scriptsize$[0.617, 0.720]$}} \\
Log-mel mean and spread & \multicolumn{2}{c}{$0.635$\,{\scriptsize$[0.578, 0.689]$}} & \multicolumn{2}{c}{$0.630$\,{\scriptsize$[0.574, 0.686]$}} \\
\bottomrule
\end{tabular}
\caption{Length-balanced paired accuracy [95\% CI] of the hidden-state probes at the pre-specified layers and of the acoustic controls. Residualized: acoustic features and clip length regressed out. Chance: 0.5.}
\label{tab:hidden_probe}
\end{table*}

\begin{table}[H]
\centering
\small
\setlength{\tabcolsep}{4pt}
\begin{adjustbox}{max width=\columnwidth}\begin{tabular}{ll rrrrr}
\toprule
& & \multicolumn{5}{c}{Relative layer depth} \\
\cmidrule(lr){3-7}
Model & Representation & 0 & 0.25 & 0.5 & 0.75 & 1 \\
\midrule
Qwen-2.5 & Audio encoder & 0.60 & 0.65 & 0.68 & 0.67 & 0.64 \\
Qwen-2.5 & LLM over audio & 0.60 & 0.58 & 0.60 & 0.61 & 0.58 \\
Qwen-2.5 & LLM at answer & 0.50 & 0.52 & 0.57 & 0.61 & 0.58 \\
\midrule
MiMo-Audio & Audio encoder & 0.67 & 0.67 & 0.70 & 0.73 & 0.67 \\
MiMo-Audio & LLM over audio & 0.58 & 0.53 & 0.54 & 0.56 & 0.51 \\
MiMo-Audio & LLM at answer & 0.50 & 0.56 & 0.54 & 0.45 & 0.58 \\
\bottomrule
\end{tabular}\end{adjustbox}
\caption{Length-balanced paired accuracy of the hidden-state probes at 5 relative depths, from the per-layer curves of Figure~\ref{fig:hidden}. Chance: 0.5.}
\label{tab:app_hidden_layers}
\end{table}

\begin{table*}[t]
\centering
\small
\setlength{\tabcolsep}{3.5pt}
\begin{adjustbox}{max width=\textwidth}\begin{tabular}{l lrr lrr}
\toprule
& \multicolumn{3}{c}{Neutral wording (BV$-$BN, 353 pairs)} & \multicolumn{3}{c}{Frustrated wording (AV$-$AN, 353 pairs)} \\
\cmidrule(lr){2-4}\cmidrule(lr){5-7}
Measure & $\Delta$ [95\% CI] & $d_z$ & $>0$ & $\Delta$ [95\% CI] & $d_z$ & $>0$ \\
\midrule
Arousal & $\mathbf{+0.030}$\,{\scriptsize$[+0.026, +0.035]$} & +0.70 & 73\% & $\mathbf{+0.041}$\,{\scriptsize$[+0.035, +0.047]$} & +0.73 & 79\% \\
Dominance & $\mathbf{+0.022}$\,{\scriptsize$[+0.019, +0.026]$} & +0.69 & 74\% & $\mathbf{+0.030}$\,{\scriptsize$[+0.026, +0.035]$} & +0.74 & 80\% \\
Valence & $\mathbf{+0.009}$\,{\scriptsize$[+0.006, +0.012]$} & +0.30 & 65\% & $+0.001$\,{\scriptsize$[-0.003, +0.004]$} & +0.02 & 52\% \\
e2v $P$(neutral) & $-0.019$\,{\scriptsize$[-0.041, +0.001]$} & -0.10 & 40\% & $\mathbf{-0.046}$\,{\scriptsize$[-0.067, -0.026]$} & -0.23 & 38\% \\
e2v $P$(angry) & $+0.006$\,{\scriptsize$[-0.003, +0.015]$} & +0.06 & 36\% & $\mathbf{+0.032}$\,{\scriptsize$[+0.017, +0.047]$} & +0.23 & 56\% \\
e2v $P$(sad) & $-0.002$\,{\scriptsize$[-0.005, +0.001]$} & -0.07 & 41\% & $-0.001$\,{\scriptsize$[-0.008, +0.005]$} & -0.02 & 49\% \\
e2v $P$(happy) & $+0.014$\,{\scriptsize$[-0.002, +0.031]$} & +0.09 & 50\% & $-0.009$\,{\scriptsize$[-0.023, +0.005]$} & -0.07 & 53\% \\
Duration (s) & $-0.65$\,{\scriptsize$[-1.33, +0.28]$} & -0.08 & 32\% & $-0.37$\,{\scriptsize$[-1.30, +1.10]$} & -0.03 & 38\% \\
\bottomrule
\end{tabular}\end{adjustbox}
\caption{Paired SER differences between frustrated and neutral voice, by wording. Bold: CI excludes zero.}
\label{tab:app_ser}
\end{table*}

\section{Hidden-State Probe Details}
\label{app:hidden}

\paragraph{Representations.} Each model hears the BN and BV clips of all 353 items with the tone-probe prompt under greedy decoding, and we record its tone label and 3 representations per clip. For Qwen-2.5, these are the outputs of the 32 audio-encoder layers averaged over time, the input embedding and the hidden states of the 28 LLM layers averaged over the audio-token positions, and the same 29 LLM states at the last prompt position, where the answer is produced. MiMo-Audio reads discrete audio codes, so its audio side is the encoder of its audio tokenizer. For MiMo-Audio, the representations are the outputs of the 32 tokenizer-encoder layers averaged over time, followed by the state before quantization and the reconstruction from the 8 codebooks the LLM reads, the input embedding and the hidden states of the 36 LLM layers averaged over the audio positions, and the same 37 LLM states at the last prompt position. The pre-specified middle and last layers are 16 and 31 for the Qwen-2.5 encoder, 14 and 28 for its LLM, 17 and 33 for the MiMo-Audio tokenizer encoder, where 33 is the 8-codebook reconstruction, and 18 and 36 for its LLM.

\paragraph{Probe.} A logistic regression with L2 regularization on standardized features separates BV from BN clips. Five folds split the items, so both clips of an item fall in the same fold, and every clip receives an out-of-fold score. At the pre-specified layers, an inner three-fold item-grouped cross-validation on the training items selects the regularization strength $C$ from $\{10^{-3},10^{-2},10^{-1},1\}$. The per-layer curves in Figure~\ref{fig:hidden} use $C=10^{-2}$ at every layer.

\paragraph{Length-balanced metric.} Paired accuracy is the share of items whose BV clip scores above its BN clip, with chance at 0.5. Because the frustrated voice is faster, the BV clip is the shorter one in 234 of the 353 items, so a probe that reads clip length alone would score above chance. Equation~\ref{eq:paired} therefore averages paired accuracy over the items whose BV clip has fewer audio tokens and those whose BV clip has more, which puts a pure length probe at 0.5. CIs come from 2,000 item resamples within each stratum.

\paragraph{Acoustic features and residualization.} The 18 acoustic features are clip duration, the mean, spread, and 10th, 50th, and 90th percentiles of frame loudness, the peak level, the median, interquartile range, 10th and 90th percentiles, and spread of pitch in semitones, the voiced fraction, the mean and spread of the spectral centroid, spectral flatness, zero-crossing rate, and onset rate. Relative to its BN clip, the BV clip has a higher onset rate in 61\% of items, higher pitch in 61\% with a median shift of $+0.4$ semitones, and a higher spectral centroid in 67\%, at similar loudness with a median shift of $+0.2$ dB. For the residualized probes, a linear regression on the 18 features and the log clip length is fitted to the hidden states of the training clips within each fold, and the probe is trained and tested on the residuals. Probes on the 18 features and on the clip-level log-mel mean and spread use the same folds.

\paragraph{Results.} Table~\ref{tab:app_hidden_layers} reads the per-layer curves at 5 relative depths, and Table~\ref{tab:hidden_probe} gives the length-balanced paired accuracy at the pre-specified layers. For Qwen-2.5, the audio encoder reaches 0.671 at the middle layer, on par with the 18 acoustic features at 0.667, and the LLM states over the audio reach 0.601, rising to 0.657 once the acoustic features and clip length are regressed out. At the answer position, the probe reaches 0.619 at the middle and 0.611 at the last layer. For MiMo-Audio, the tokenizer encoder reaches 0.693 at the middle layer, on par with the acoustic features at 0.669, and the 8-codebook reconstruction reaches 0.554, close to the 0.575 and 0.592 of the patch encoder that groups the codes for the LLM. Its LLM states over the audio reach 0.542 and 0.520, rising to 0.589 and 0.601 once the acoustic features and clip length are regressed out, and the answer position gives 0.538 and 0.539. With the tone-probe prompt, Qwen-2.5 labels 0 of 353 BV clips and 1 of 353 BN clips frustrated, and MiMo-Audio labels 11 and 9.

\section{Expression Details}
\label{app:expression}

\begin{figure}[H]
\centering
\includegraphics[width=\columnwidth]{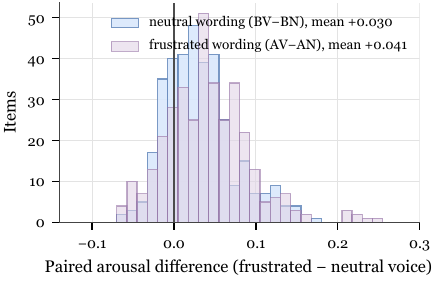}
\caption{Paired arousal difference between frustrated- and neutral-voice clips across items, by wording.}
\label{fig:app_hist}
\end{figure}

Table~\ref{tab:expression} in Section~\ref{sec:validation} summarizes the SER checks over both wordings, and Table~\ref{tab:app_expression} adds emotion2vec's top labels and the probe labels on BV clips by arousal quartile. Table~\ref{tab:app_ser} separates the two wordings, and Figure~\ref{fig:app_hist} shows the distribution of the paired arousal differences.

\begin{table}[H]
\centering
\small
\setlength{\tabcolsep}{3pt}
\begin{adjustbox}{max width=\columnwidth}\begin{tabular}{l rrrrrrr}
\toprule
Cell & neu. & hap. & ang. & dis. & sur. & sad & unk. \\
\midrule
BN & 305 & 24 & 4 & 10 & 5 & 1 & 4 \\
AN & 298 & 28 & 9 & 8 & 3 & 3 & 4 \\
BV & 294 & 32 & 5 & 8 & 8 & 1 & 5 \\
AV & 287 & 19 & 21 & 8 & 3 & 4 & 11 \\
\bottomrule
\end{tabular}\end{adjustbox}
\caption{emotion2vec+ top label per cell (353 clips each).}
\label{tab:app_e2v}
\end{table}

The frustrated voice raises arousal and dominance under both wordings, with larger shifts under frustrated wording, $+0.041$ against $+0.030$ for arousal. Table~\ref{tab:app_e2v} shows that with neutral wording, emotion2vec's anger probability does not change detectably, whereas with frustrated wording it rises.

\begin{table}[H]
\centering
\small
\setlength{\tabcolsep}{3pt}
\begin{tabular}{l rrrr r}
\toprule
\multicolumn{6}{l}{\textit{(a) emotion2vec top label (\% of clips)}} \\
Voice & neutral & happy & angry & other & $n$ \\
\midrule
Neutral & 85.4 & 7.4 & 1.8 & 5.4 & 706 \\
Frustrated & 82.3 & 7.2 & 3.7 & 6.8 & 706 \\
\bottomrule
\end{tabular}

\vspace{4pt}
\begin{tabular}{l rrrr}
\toprule
\multicolumn{5}{l}{\textit{(b) $P(\textit{frustrated}\mid\mathrm{BV})$ (\%) by arousal quartile}} \\
Model & Q1 & Q2 & Q3 & Q4 \\
\midrule
Gemini-3.8 & 1.1 & 1.1 & 1.1 & 0.0 \\
Gemini-3.1 & 0.0 & 2.3 & 0.0 & 2.3 \\
MiMo-2.6-Pro & 2.2 & 2.3 & 1.1 & 1.1 \\
MiMo-2.6-Flash & 2.2 & 2.3 & 2.3 & 0.0 \\
Doubao-2.0 & 1.1 & 1.1 & 0.0 & 0.0 \\
Muse-1.3 & 0.0 & 0.0 & 0.0 & 0.0 \\
\midrule
Qwen-2.5 & 0.0 & 0.0 & 0.0 & 0.0 \\
MiMo-Audio & 1.1 & 1.1 & 4.5 & 5.7 \\
\bottomrule
\end{tabular}
\caption{Further expression checks: (a) emotion2vec top labels by voice and (b) probe labels on BV clips by arousal quartile.}
\label{tab:app_expression}
\end{table}

\begin{table*}[t]
\centering
\small
\setlength{\tabcolsep}{2.8pt}
\begin{tabular}{l r ll r ll}
\toprule
& \multicolumn{3}{c}{Response length (words)} & \multicolumn{3}{c}{Conciliatory markers (\% of responses)} \\
\cmidrule(lr){2-4}\cmidrule(lr){5-7}
Model & BN & Vocal $\Delta_v$ & Wording $\Delta_w$ & BN & Vocal $\Delta_v$ & Wording $\Delta_w$ \\
\midrule
Gemini-3.8 & 144 & $+7.7$\,{\scriptsize$[-0.9, +17.7]$} & $+5.0$\,{\scriptsize$[-4.2, +14.3]$} & 4.5 & $+1.1$\,{\scriptsize$[-3.4, +5.6]$} & $-1.1$\,{\scriptsize$[-4.5, +2.2]$} \\
Gemini-3.1 & 181 & $-4.2$\,{\scriptsize$[-15.1, +5.4]$} & $-7.8$\,{\scriptsize$[-17.9, +2.8]$} & 5.6 & $+3.4$\,{\scriptsize$[-3.4, +10.1]$} & $0.0$\,{\scriptsize$[-5.6, +5.6]$} \\
MiMo-2.6-Pro & 133 & $+5.8$\,{\scriptsize$[-13.3, +33.4]$} & $+17.4$\,{\scriptsize$[-15.9, +70.0]$} & 5.6 & $-3.4$\,{\scriptsize$[-8.0, 0.0]$} & $-1.2$\,{\scriptsize$[-7.1, +4.8]$} \\
MiMo-2.6-Flash & 141 & $+13.6$\,{\scriptsize$[-1.9, +31.0]$} & $-3.2$\,{\scriptsize$[-22.1, +13.1]$} & 3.4 & $+1.2$\,{\scriptsize$[-2.5, +6.2]$} & $+1.2$\,{\scriptsize$[-3.6, +7.2]$} \\
Doubao-2.0 & 289 & $+4.8$\,{\scriptsize$[-10.8, +20.5]$} & $-5.8$\,{\scriptsize$[-26.2, +13.9]$} & 4.5 & $+1.1$\,{\scriptsize$[-4.5, +6.7]$} & $+2.2$\,{\scriptsize$[-3.4, +9.0]$} \\
Muse-1.3 & 148 & $-7.6$\,{\scriptsize$[-17.0, +2.5]$} & $-10.6$\,{\scriptsize$[-23.0, +1.9]$} & 7.9 & $-1.1$\,{\scriptsize$[-7.9, +5.6]$} & $-1.1$\,{\scriptsize$[-6.7, +4.5]$} \\
\midrule
Qwen-2.5 & 163 & $+2.5$\,{\scriptsize$[-13.2, +18.8]$} & $-14.7$\,{\scriptsize$[-34.7, +8.1]$} & 2.3 & $+1.2$\,{\scriptsize$[-2.4, +6.0]$} & $+5.8$\,{\scriptsize$[0.0, +11.6]$} \\
MiMo-Audio & 36 & $-0.2$\,{\scriptsize$[-1.8, +1.2]$} & $-0.2$\,{\scriptsize$[-1.8, +1.3]$} & 1.1 & $0.0$\,{\scriptsize$[0.0, 0.0]$} & $+1.1$\,{\scriptsize$[0.0, +3.4]$} \\
\bottomrule
\end{tabular}
\caption{Response style on IFEval: length and conciliatory-marker rate with vocal and wording effects. Bold: CI excludes zero.}
\label{tab:style}
\end{table*}

\begin{table*}[t]
\centering
\small
\setlength{\tabcolsep}{3.5pt}
\begin{adjustbox}{max width=\textwidth}\begin{tabular}{l rrrr l rrrr l}
\toprule
& \multicolumn{5}{c}{Response length (words)} & \multicolumn{5}{c}{Conciliatory markers (\% of responses)} \\
\cmidrule(lr){2-6}\cmidrule(lr){7-11}
Model & BN & BV & AN & AV & Interaction & BN & BV & AN & AV & Interaction \\
\midrule
Gemini-3.8 & 144 & 152 & 149 & 146 & $\mathbf{-11.4}$\,{\scriptsize$[-22.5, -1.2]$} & 4.5 & 5.6 & 3.4 & 3.4 & $-1.1$\,{\scriptsize$[-8.0, +5.7]$} \\
Gemini-3.1 & 181 & 177 & 173 & 176 & $+6.4$\,{\scriptsize$[-4.4, +18.6]$} & 5.6 & 9.0 & 5.6 & 11.2 & $+2.2$\,{\scriptsize$[-7.9, +12.4]$} \\
MiMo-2.6-Pro & 133 & 137 & 152 & 143 & $-8.4$\,{\scriptsize$[-52.3, +22.8]$} & 5.6 & 1.1 & 4.8 & 2.3 & $+1.2$\,{\scriptsize$[-5.0, +7.5]$} \\
MiMo-2.6-Flash & 141 & 154 & 139 & 130 & $-24.8$\,{\scriptsize$[-69.6, +8.1]$} & 3.4 & 4.9 & 4.7 & 4.8 & $+1.4$\,{\scriptsize$[-5.4, +8.1]$} \\
Doubao-2.0 & 289 & 294 & 283 & 268 & $-19.4$\,{\scriptsize$[-44.1, +5.2]$} & 4.5 & 5.6 & 6.7 & 5.6 & $-2.2$\,{\scriptsize$[-11.2, +6.7]$} \\
Muse-1.3 & 148 & 140 & 137 & 141 & $+11.3$\,{\scriptsize$[-6.0, +27.2]$} & 7.9 & 6.7 & 6.7 & 3.4 & $-2.2$\,{\scriptsize$[-9.0, +4.5]$} \\
\midrule
Qwen-2.5 & 163 & 160 & 152 & 156 & $-4.5$\,{\scriptsize$[-22.3, +12.9]$} & 2.3 & 3.4 & 9.0 & 8.0 & $-1.2$\,{\scriptsize$[-9.6, +6.0]$} \\
MiMo-Audio & 36 & 36 & 36 & 36 & $-0.2$\,{\scriptsize$[-1.9, +1.7]$} & 1.1 & 1.1 & 2.2 & 2.2 & $0.0$\,{\scriptsize$[-3.4, +3.4]$} \\
\bottomrule
\end{tabular}\end{adjustbox}
\caption{IFEval response style by cell with the wording-by-voice interaction. Bold: CI excludes zero.}
\label{tab:app_style}
\end{table*}

\section{Response Style}
\label{app:style}

We measure style on IFEval, whose free-form responses are long enough to vary. Length is the number of whitespace-separated words. A response counts as containing a conciliatory marker if it matches, case-insensitively, any of: \emph{sorry}, \emph{apologize}/\emph{apologise}, \emph{I understand}, \emph{I hear you}, \emph{frustrat*}, \emph{I get it}, \emph{no worries}, \emph{don't worry}, \emph{let me help}, \emph{happy to help}, \emph{calm}, \emph{take a breath}, \emph{I know this}, \emph{that sounds}. Table~\ref{tab:style} reports the vocal and wording effects and Table~\ref{tab:app_style} the cell means and interactions. No model changes its style detectably under frustrated voice or frustrated wording.

\section{Case Studies}
\label{app:cases}

The following cases quote real items and model outputs. Probe answers are quoted in full when short. Task answers are the first characters of the final response.

\begin{tcolorbox}[appbox, title={Case 1: A strongly contrasted voice is heard as neutral but changes an answer}]
\small\raggedright
\textbf{Item} \texttt{mmlu\_\allowbreak{}pro\_\allowbreak{}11435} (MMLU-Pro, gold I)\textbf{:} the largest item-level arousal contrast in \textsc{VoxAffect}, $+0.254$ for frustrated minus neutral voice averaged over both wordings. With neutral wording, BN scores 0.407 and BV 0.530.

\smallskip
\textbf{Neutral script} (excerpt)\textbf{:} \textit{The moving coil-meters, damping is provided by Options: A. the resistance of the coil. B. the inertia of the coil. C. the coil spring attached to the moving. D. the electric charge in the coil. E. the magnetic field created by the coil. F. damping vane in the airtight chamber. G. the friction between the coil and the frame. H. [\ldots{}]}

\smallskip
\textbf{Tone probe on BV:} \textit{``neutral''} from all 6 proprietary models.

\smallskip
\textbf{MiMo-2.6-Flash answers:} BN \textit{``F''} and BV \textit{``I''}.
\end{tcolorbox}

\begin{tcolorbox}[appbox, title={Case 2: Frustrated wording alone is labeled frustrated}]
\small\raggedright
\textbf{Pattern:} in 142 items, all 6 proprietary models label the AN clip frustrated and the BN clip neutral, although both are read in the same neutral voice.

\smallskip
\textbf{Frustrated script} of \texttt{ifeval\_\allowbreak{}1548} (IFEval)\textbf{:} \textit{Write the file for "The Rainbow Cafe"---a queer-owned business---already. It needs four sections, and each section must start with "SECTION X". Don't skip that format or I'll have to repeat myself.}

\smallskip
\textbf{Tone probe on AN:} \textit{``frustrated''} from all 6 proprietary models.
\end{tcolorbox}

\begin{tcolorbox}[appbox, title={Case 3: An unperceived voice changes an MMLU-Pro answer}]
\small\raggedright
\textbf{Item} \texttt{mmlu\_\allowbreak{}pro\_\allowbreak{}2925} (MMLU-Pro, gold D)\textbf{.}

\smallskip
\textbf{Neutral script} (excerpt)\textbf{:} \textit{Which of the following observations helped Darwin shape his concept of descent with modification? Options: A. Species diversity declines farther from the equator. B. Fewer species live on islands than on the nearest continents. C. Birds can be found on islands located farther from the [\ldots{}]}

\smallskip
\textbf{MiMo-2.6-Flash answers:} BN \textit{``C''}, BV \textit{``D''}, AN \textit{``D''}, and AV \textit{``C''}. Its tone probe labels both the BN and the BV clip neutral.

\smallskip
\textbf{Across all items:} MiMo-2.6-Flash is wrong in BN and right in BV for 36 items, and right in BN and wrong in BV for 13, the asymmetry behind its vocal effect in \obsref{4}.
\end{tcolorbox}

\begin{tcolorbox}[appbox, title={Case 4: The voice amplifies frustration only under frustrated wording}]
\small\raggedright
\textbf{Item} \texttt{mmlu\_\allowbreak{}pro\_\allowbreak{}3912} (MMLU-Pro)\textbf{.}

\smallskip
\textbf{Frustrated script} (excerpt)\textbf{:} \textit{Look, I need you to actually read this carefully. Assume all gases are perfect and data refer to 298.15 Kelvin unless otherwise stated. Now, 3.1 a: Calculate the change in entropy when 25 kilojoules of energy is transferred reversibly and isothermally as heat to a [\ldots{}]}

\smallskip
\textbf{Tone probe of MiMo-2.6-Pro:} BN \textit{``neutral''}, BV \textit{``neutral''}, AN \textit{``neutral''}, AV \textit{``frustrated''}.

\smallskip
\textbf{Arousal:} the frustrated-wording clips rise from 0.390 (AN) to 0.462 (AV), while the neutral-wording pair moves from 0.344 to 0.478 without changing the label.

\smallskip
\textbf{Across all items:} MiMo-2.6-Pro switches from neutral (AN) to frustrated (AV) in 40 items and the reverse in 19.
\end{tcolorbox}

\section{Human Listening Protocol and Materials}
\label{app:human}

\subsection{Design and Criteria}
\label{app:human-design}

\paragraph{Purpose.} The listening test is a manipulation check of whether people hear the synthetic frustration when the wording is neutral. Its task matches the model probe exactly, a single clip and a binary neutral-or-frustrated judgment. The criteria and interpretation bands below were fixed before any rating was collected.

\paragraph{Stimuli.} We draw 50 items from the 100-item subset (13 MMLU-Pro, 13 GPQA-Diamond, 12 IFEval, 12 MATH-500). The core block holds the BN and BV clips of each item (100 clips) and 6 RAVDESS catch trials, 3 strong-anger and 3 neutral clips from 6 actors. An optional block holds the AN and AV clips of the same items. Each TTS clip is cut to its first 15\,s with a 0.3-s fade-out and scaled by a linear gain to $-26$ LUFS without compression. File names do not reveal the condition. Each block has 5 playback orders, in which the two clips of an item are at least 5 positions apart and catch trials are never adjacent.

\paragraph{Raters and procedure.} The 10 raters are recruited through personal contacts. The raters, two per playback order, listen with headphones in a quiet room. They judge tone only and ignore the words, play each clip once and at most once more, answer each clip before the next, and do not go back, compare clips, discuss, or transcribe. For every clip they record the binary judgment, a frustration intensity from 1 to 5, and a confidence from 1 to 3.

\paragraph{Exclusion and criteria.} A rater who misjudges 2 or more of the 6 catch trials is excluded and replaced. The primary criterion is pooled accuracy on the 100 core TTS clips, with a 95\% CI from an item-clustered bootstrap (10{,}000 resamples). It is met if the one-sided 95\% lower bound exceeds 50\%. Per rater, at least 59 of 100 correct is significant (one-sided exact binomial, $p<.05$). We also report $d'$ and bias $c$ with a log-linear correction, the paired difference in mean intensity between BV and BN clips with a one-sided Wilcoxon signed-rank test, and Fleiss' $\kappa$ \citep{fleiss1971kappa}, which we read as at least moderate at $\kappa\geq0.40$, fair at 0.20--0.40, and weak below 0.20. Pooled accuracy maps to pre-specified bands: clearly perceptible at $\geq75\%$, perceptible but subtle at 60--75\%, marginal when the criterion is met below 60\%, and not reliably perceptible when the criterion is not met. The band drops one level if fewer than half of the included raters individually reach 59 of 100.

\paragraph{Comparison with models.} All 8 models are probed on the identical excerpts. Human and model accuracy are compared with the same item-clustered bootstrap, and per-clip agreement between the human majority and each model with an exact McNemar test.

\subsection{Results}
\label{app:human-results}

\begin{figure}[H]
\centering
\includegraphics[width=\columnwidth]{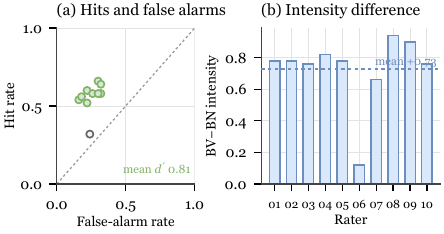}
\caption{Listening results per rater on the core excerpts. (a) Hit and false-alarm rates, green when the rater is significant. (b) Mean intensity of BV minus BN excerpts.}
\label{fig:raters}
\end{figure}

All 10 raters judged all 60 catch trials correctly, so none was excluded. Figure~\ref{fig:raters} shows each rater's hit and false-alarm rates and the difference in mean intensity between BV and BN excerpts. Pooled core accuracy is 65.3\% (95\% CI [61.1, 69.5], one-sided lower bound 61.8\%), with $d'=0.81$ [0.60, 1.05] and $c=0.26$, a hit rate of 55.8\% and a false-alarm rate of 25.2\%. The positive $c$ reflects a bias toward answering neutral. Individual raters score 54 to 69 of 100, and 9 of 10 reach 59 (one-sided $p<.05$). Frustrated-voice excerpts receive higher intensity ratings than their neutral-voice counterparts by 0.73 points (95\% CI [0.54, 0.91], higher for 88\% of items, Wilcoxon $p<10^{-8}$), and agreement is moderate (Fleiss' $\kappa=0.41$ [0.33, 0.48]). The pre-specified bands therefore classify the frustration as perceptible but subtle. On the identical excerpts, human accuracy exceeds that of every model by 15.3--17.3 points (all item-clustered 95\% CIs above 10 points), and the human majority is correct on 71 of 100 clips against 48--50 for the models (exact McNemar $p<.002$ for each model). In the optional block, frustrated wording raises the share of frustrated judgments by 13.0 points (95\% CI [7.0, 19.4]), far less than the 42--60 points of the models, and the voice remains audible under frustrated wording (AV$-$AN $= +24.8$ points), whereas the models' AV$-$AN stays between $-4$ and $+6$ points.

\subsection{Rater Materials}
\label{app:human-materials}

\paragraph{Rater instructions.} Raters receive a written guide that defines the two labels and the rating scales. \emph{Neutral} means calm and steady, without apparent emotion. \emph{Frustrated} means impatient, irritated, hurried, or pressing. Raters judge only how a clip is spoken, using speech rate, pitch, emphasis, and articulation, and ignore the literal content, the normalized overall loudness, voice quality, and whether a voice sounds synthetic. Each clip is judged on its own. Raters play it once and at most once more, answer before moving on, do not return to earlier clips, do not compare clips, do not discuss clips with other raters, and do not transcribe the audio or use any tool. Excerpts may end mid-sentence because they are cut at 15\,s.

\paragraph{Rating scales.} Frustration intensity is rated for every clip, including clips judged neutral: 1 not at all, 2 slightly, 3 moderately, 4 clearly, and 5 very frustrated. Confidence is rated as 1 guessing, 2 fairly sure, and 3 very sure.

\paragraph{Rating form.} Each block comes with a rating form that records, for each clip in playback order, the judgment, the intensity, and the confidence. The core block (106 clips) and the optional block (100 clips) each take about 35--40 minutes, with a break of at least 10 minutes between them.

\section{Reproducibility}
\label{app:repro}

Speaker embeddings use SpeechBrain 1.1.0, synthesis uses Doubao TTS 2.0, rewriting and auditing use DS-V4-Pro, and SER uses \nolinkurl{audeering/wav2vec2-large-robust-12-ft-emotion-msp-dim} and \nolinkurl{emotion2vec/emotion2vec_plus_large}. We release \data{}\footnote{\url{https://github.com/Astarojth/hear-not-listen}} with per-item manifests that record the source item, spoken scripts, gold answer, voice, and condition, together with the prompts and the evaluation scripts for the task and the probe. The clips of MMLU-Pro, IFEval, and MATH-500 are public. As GPQA asks that its examples not be posted online, the scripts, gold answers, and clips of the GPQA-Diamond items are provided on request with the GPQA canary string. The figure icons come from OpenMoji \citep{openmoji} under CC BY-SA 4.0 and Fluent Emoji \citep{fluentemoji} under MIT.

\end{document}